\documentclass[pre,twocolumn,a4paper,showkeys,showpacs]{revtex4-1}
\usepackage[dvipdfmx]{graphicx}
\usepackage{bmpsize}
\usepackage{calc}
\usepackage{amsmath}
\usepackage{physics}
\usepackage{bm}         
\usepackage{mathrsfs}  
\usepackage{amssymb}
\usepackage{comment}
\usepackage{cases}
\usepackage{here}
\usepackage{array}
\usepackage{hyperref}
\usepackage{xr}
\usepackage{xcolor}
\usepackage{tabularx}
\newcolumntype{C}{>{\centering\arraybackslash}X}
\newcolumntype{L}{>{\raggedright\arraybackslash}X}
\newcolumntype{R}{>{\raggedleft\arraybackslash}X}
\usepackage{ulem}

\usepackage{etoolbox}
\makeatletter
\newcounter{firstbib}
\apptocmd{\thebibliography}{%
  \setcounter{NAT@ctr}{\value{firstbib}}%
}{}{}
\AtEndEnvironment{thebibliography}{%
  \setcounter{firstbib}{\value{NAT@ctr}}%
}
\makeatother

\usepackage{lipsum}

\newcommand{\leri}[1]{\left({#1}\right)}
\newcommand{\sgn}[1]{\mathrm{sgn}\leri{{#1}}}

\begin{document}

\title{
Optic flow in the visual system determines interaction force and collective motion
}
\author{Susumu Ito$^1$}
\thanks{susumu.ito.b4@tohoku.ac.jp}
\author{Nariya Uchida$^2$}
\affiliation{$^1$Department of Biomedical Engineering, Tohoku University, Sendai, 980-8579, Japan\\
$^2$Department of Physics, Tohoku University, Sendai, 980-8578, Japan}
\date{\today} 

\begin{abstract}
Collective motion in animal groups relies on visual cues, 
yet how individual sensory inputs translate into group-level dynamics remains unclear. 
Traditional models often assume pre-defined ``3A" rules -- avoidance, attraction, and alignment -- 
which require agents to directly access the exact positions and velocities of their neighbors. 
Here, we propose a minimal model based on fundamental visual processing shared across species: 
detection of image size and optic flow. 
We show that the collective patterns, such as vortex and polarized school, 
emerge spontaneously from the competition between 
image-size-driven avoidance and optic-flow-driven approach. 
We find that alignment does not require a pre-assumed velocity-matching rule; 
instead, it arises from the acceleration of an agent toward regions of strong optic flow. 
The model shows scale-free velocity correlations and topological visual features observed in animal groups. 
Our results indicate that the phenomenological 3A forces are not irreducible physical rules, 
but rather emergent outcomes of basic sensorimotor feedback. 
This framework provides a physical perspective for linking sensory inputs to macroscopic collective behaviors.
\end{abstract}

\maketitle

Collective motion is ubiquitous among various organisms~\cite{Vicsek2012}.
In organisms with eyes, visual information plays a crucial role~\cite{Peshkin2013,Sridhar2021}.
Observations of collective behavior characterize the features of visual interactions based on 
the distribution of force between individuals~\cite{Katz2011,Herbert2011,Calovi2018},
and the correlations in their spatial positions~\cite{Ballerini2008,Rosenthal2015}
and relative velocity~\cite{Sridhar2023,Puy2024}.
However, it is unclear how such effective forces and correlations emerge from sensory inputs.

Recent experiments on the visual circuits show the relation between neural activity and individual responses.
For example, in well-studied fish cases, specific visual stimuli such as image size and motion 
relate to behaviors: avoidance~\cite{Dunn2016,Bhattacharyya2017},
approach and turning to visual stimuli~\cite{Kappel2022,Dragomir2020}, 
mediated by retinotectal neural circuits.
Despite this progress, it remains technically difficult to monitor 
the dynamics of these circuits in individuals during collective motion.

Therefore, constructing a model that links individual sensory processing to
collective motion is a crucial challenge.
Traditionally, collective motion has been addressed using phenomenological 
``3A" (avoidance, attraction, and alignment) models~\cite{Couzin2002,Hemelrijk2012}, 
which require agents to directly access both 
the exact spatial positions and relative velocities of neighbors
within a metric or topological range~\cite{DOrsogna2006,Gautrais2012,Calovi2014,Ito2022}.
To ground 3A rules, several models introduce visual information in various ways:
using the edges of the visual image~\cite{Pearce2014,Bastien2020},
stochastic decision-making~\cite{Collignon2016},
energy minimization~\cite{Sridhar2021,Heins2024},
and the motion of a neighbor on the line of sight~\cite{Ito2024}.
These models successfully infer neighbor positions from image size, 
while they assume that relative velocity can be directly perceived.
To capture direct relative motion information,
a few models introduce optic flow 
which is the apparent angular velocity of images in the visual field~\cite{Lemasson2009,Castro2024}.
In these studies, optic flow is employed primarily to reconstruct the relative velocity of neighbors,
and then is used for velocity matching as alignment in 3A models.
In particular, Castro \textit{et al.}~\cite{Castro2024} combined optic flow 
with an apparent distance inferred from the horizontal angular size of visible images 
to construct a visual counterpart of distance-dependent attraction and alignment. 
However, when images of multiple neighbors overlap or occlude one another, 
the distance to each individual cannot in general be uniquely inferred. 

In this study, 
we propose a biologically inspired visual interaction model
incorporating optic flow. 
Our formulation directly uses image size and optic flow without reconstructing neighbor distance or relative velocity.
Recognizing the inherent complexity of visual systems, 
we formulate this as a minimal yet illustrative framework among various possible mechanisms.
The model is based on visual processing mechanisms found across diverse species, 
from insects to vertebrates, and does not assume 3A rules.
We systematically derive these interactions by spatial mirror symmetry.

Through this formulation, the equilibrium distance between individuals spontaneously emerges 
from the competition between image-size-driven avoidance and optic-flow-driven approach,
instead of a pre-defined static equilibrium distance.
Although individuals react only to the visual stimuli from near neighbors, 
the diverse collective motion exhibits intermittency of the pattern change on the long time scale
and scale-free correlations in velocity fluctuations.
Rather than assuming direct velocity-matching rules, we demonstrate that
the alignment emerges from acceleration toward the region of strong optic flow.
This work offers a unifying perspective: 
individuals react to visual stimuli, and the widely accepted 3A forces
are explained not as pre-existing irreducible physical rules, 
but as emergent outcomes of these intrinsic sensory responses.

\section*{Construction of model from visual system}

Our model is built upon fundamental visual features conserved across species for image size and optic flow. 
First, collision avoidance is triggered
by the projected image size~\cite{Harpaz2021,Dewell2023}.
Second, the visual system detects the optic flow
by matching a delayed signal from one spatial location 
with the instantaneous signal from a neighboring location.
T4/T5 cells in insects~\cite{Maisak2013} and 
direction selective retinal ganglion cells (DS-RGCs) in vertebrates~\cite{Yonehara2011,Nikolaou2012}
mediate this function.
The Hassenstein-Reichardt correlator (HRC)~\cite{Egelhaaf1987}
describes this mechanism~\cite{Kim2014}.
Finally, the neural circuits of diverse animal species possess the mechanisms of inhibition 
which increases selective sensitivity to weak stimuli from small images
for foraging and recognizing conspecifics~\cite{Wiederman2008,Isa2021}.
For example, superficial interneurons (SINs) in fish mediate this role~\cite{Preuss2014,Yin2019}
through exponential inhibition~\cite{Dunn2016}. 

\begin{figure*}[!t]
\centering
\includegraphics[width=\linewidth]{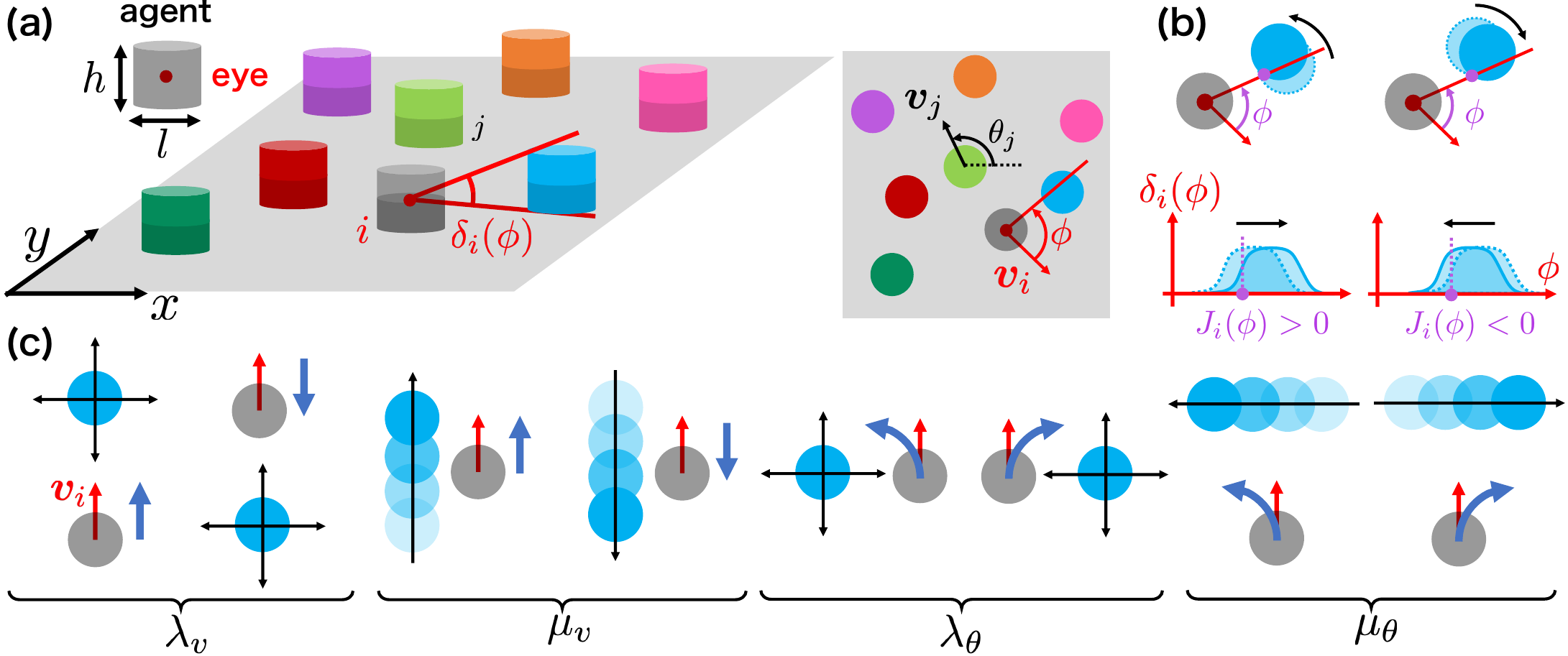}
\caption{
Schematic illustration of our model.
(a) Cylinder-shaped agents with diameter $l$, height $h$, and a monoeye at center move on a $x$-$y$ plane.
The right figure is a top view of the left figure.
The angle of direction of motion $\theta$ is measured from $x$-axis.
The angle in the field of view $\phi$ is measured from the orientation of the agent along the velocity $\bm{v}$.
Both of the angles are defined in the counterclockwise direction and take values in $[-\pi,~\pi]$.
The black agent $i$ detects the vertical angular diameter $\delta_i(\phi)$ of the cyan neighbor at $\phi$.
(b) The cyan neighbor moves around the black agent.
The state from a short time ago is shown by the dashed line.
The Gaussian-blurred angular diameter $\delta_i(\phi)$ moves across the field of view. 
In this case, the value of $J_i(\phi)$ at $\phi$ is positive (left panel) and negative (right panel).
(c) Schematic illustrations of qualitative properties of Eqs.~(\ref{funcZv},~\ref{funcZt}).
The black agent reacts to the cyan neighbor, and this reaction is represented by the blue arrow.
The terms involving $\lambda_v$:
the agent accelerates toward the front neighbor 
and decelerates in response to the behind neighbor,
regardless of the sign of $J$.
The terms involving $\mu_v$:
the agent accelerates when
the lateral neighbor overtakes the agent,
and decelerates when
the lateral neighbor moves toward the back.
The terms involving $\lambda_\theta$:
the agent turns toward lateral neighbors,
regardless of the sign of $J$.
The terms involving $\mu_\theta$:
the agent turns left when the neighbor moves to counterclockwise and shows positive $J$,
turns right when the neighbor moves to clockwise and shows negative $J$.
}
\label{Fig1}
\end{figure*}

Next, we show the details of our model.
Self-propelled agents indexed by $i=1,2,...,N$ move on a plane with no boundary.
Each agent is a cylinder of diameter $l$ and height $h$
with a monoeye at center:
the position of the eye $\bm{r}_i=(x_i, y_i)$ and 
the velocity $\bm{v}_i=\dv*{\bm{r}}{t}=v_i\bm{e}_i$, 
where $v_i\geq0$ is the speed,
and $\bm{e}_i=(\cos\theta_i,\sin\theta_i)$ is the unit vector in the direction of motion
with the angle of orientation $\theta_i$ (Fig.~\ref{Fig1}(a)).

Let $\phi$ be the angle in the field of view measured from the orientation of the agent. 
We define the vertical angular diameter $\delta_i(\phi,t)$ of a neighbor
detected by the $i$th agent at the angle $\phi$ and at time $t$,
and detector signal
\begin{equation}
\label{eqJ}
J_i(\phi,t)=\delta_i\pdv{\delta_i}{\phi}{t}-\pdv{\delta_i}{\phi}\pdv{\delta_i}{t}
\end{equation}
inspired by the HRC, which extracts the cross-correlation between the spatial and temporal change of $\delta_i$
(Fig.~\ref{Fig1}(b), the Methods section ``Derivation of $J(\phi,t)$").

Here, we introduce the equation of motion of the position of the $i$th agent.
We adopt the variables ($v_i,~\theta_i$) for the equations of motion~\cite{Bastien2020,Ito2024},
and construct the interaction based on ($\delta_i,~J_i$).
The general forms without the coupling term between $\delta_i$ and $J_i$ are given by 
\begin{eqnarray}
\label{eqv}
\dv{v_i}{t}&=&C_v(v_0-v_i)+\eta_{v,i}\\\nonumber
&+&\int_{-\pi}^\pi\dd\phi\{W_v(\delta_i(\phi,t),\phi)+Z_v(J_i(\phi,t),\phi)\},
\end{eqnarray}
\begin{eqnarray}
\label{eqtheta}
\dv{\theta_i}{t}&=&\eta_{\theta,i}\\\nonumber
&+&\int_{-\pi}^\pi\dd\phi\{W_\theta(\delta_i(\phi,t),\phi)+Z_\theta(J_i(\phi,t),\phi)\},
\end{eqnarray}
where $C_v(v_0-v_i)$ is the self-propelled term with a constant $C_v$ and the steady speed $v_0$,
and $\eta_{v,i}\dd t=\sqrt{2D_v}\dd w_{v,i}$ and $\eta_{\theta,i}\dd t=\sqrt{2D_\theta}\dd w_{\theta,i}$ 
correspond to white Gaussian noises ($\dd w_{v,i}$ and $\dd w_{\theta,i}$ 
are the standard Wiener processes).
The functions $W_v,~W_\theta,~Z_v,~Z_\theta$
are formulated by considering the mirror symmetry of images
and the exponential inhibition.
We take the lower-order terms into account, and then obtain the following 
(see the Methods section ``Determination of the interaction functions").
\begin{equation}
\label{funcWv}
W_v(\delta,\phi)=-\kappa_v\delta\cos\phi,
\end{equation}
\begin{equation}
\label{funcWt}
W_\theta(\delta,\phi)=-\kappa_\theta\delta\sin\phi,
\end{equation}
\begin{equation}
\label{funcZv}
Z_v(J,\phi)=\leri{\lambda_v\abs{J}\cos\phi-\mu_vJ\sin\phi}e^{-\abs{J}/J_0},
\end{equation}
\begin{equation}
\label{funcZt}
Z_\theta(J,\phi)=\leri{\lambda_\theta\abs{J}\sin\phi+\mu_\theta J(1+\gamma\cos\phi)}e^{-\abs{J}/J_0},
\end{equation}
where $\kappa_v,~\kappa_\theta,~\lambda_v,~\lambda_\theta,~\mu_v,~\mu_\theta,~J_0$ are the positive parameters.

The qualitative properties of the interaction functions are as follows. 
To avoid collisions based on image size, 
an agent adjusts its speed to front and rear neighbors using Eq.~(\ref{funcWv}), 
and turns toward lateral neighbors using Eq.~(\ref{funcWt}). 
In contrast, 
the terms involving $\lambda_v$ in Eq.~(\ref{funcZv}) and $\lambda_\theta$ in Eq.~(\ref{funcZt}) 
cause the agent to approach its neighbors. 
This approach behavior is based on the magnitude $\abs{J}$, independent of its direction.
Furthermore, the terms involving $\mu_v$ in Eq.~(\ref{funcZv}) and $\mu_\theta$ in Eq.~(\ref{funcZt})
indicate that the agent adjusts its speed and turns to reduce $J$ (see Fig.~\ref{Fig1}(c)).

In our model, we rescaled the parameter values for nondimensionalization
by a body length (BL) $l$ and timescale of exponential inhibition $1/J_0$.
The fixed parameters are the body height $h=1$,
the coefficient of steady speed $C_v=1$,
and the steady speed $v_0=1$.
See the Methods section ``numerical methods" for details of the simulations.

\section*{Collective patterns and phase diagrams}

We have the control parameters 
$\kappa_v$, $\kappa_\theta$, $\lambda_v$, $\lambda_\theta$, $\mu_v$, $\mu_\theta$, $\gamma$, $D_v$, $D_\theta$,
and explain the effects of varying parameter values on collective patterns step by step.

\begin{figure*}[!t]
\centering
\includegraphics[width=\linewidth]{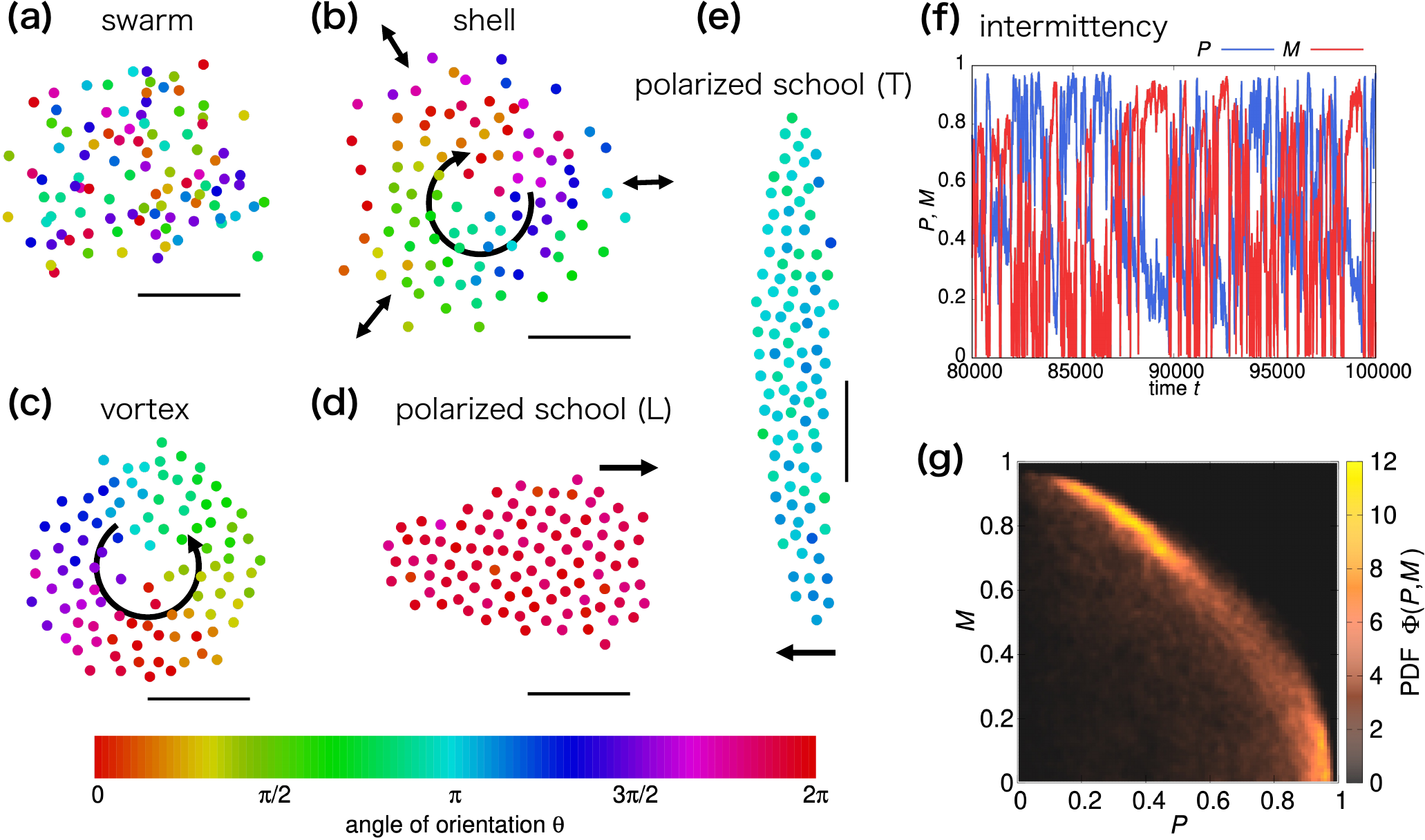}
\caption{
Collective patterns and intermittency.
(a)-(e) Snapshots of steady cluster with $N=100$,
$\kappa_\theta=1.0$, $\lambda_\theta=3.5$, $\mu_v=0.0$, $\gamma=1.0$, $D_v=0.001$, and $D_\theta=0.005$.
Each agent is represented as a circle of diameter $l=1$,
and the length of scalebar is 10.
The black arrows show the direction of motion of each agent.
The color bar below corresponds to the angle of orientation of agent.
The pattern and the parameters ($\kappa_v$, $\lambda_v$, $\mu_\theta$) are as follows.
(a) swarm: $\kappa_v=0.0$, $\lambda_v=0.0$, $\mu_\theta=0.0$.
(b) shell: $\kappa_v=3.0$, $\lambda_v=0.0$, $\mu_\theta=0.0$.
(c) vortex: $\kappa_v=1.0$, $\lambda_v=1.0$, $\mu_\theta=2.0$.
(d) polarized school (L): $\kappa_v=1.0$, $\lambda_v=3.0$, $\mu_\theta=0.0$.
(e) polarized school (T): $\kappa_v=1.0$, $\lambda_v=4.0$, $\mu_\theta=0.0$.
See Supplemental material Movies S1-S5 for the dynamics, respectively.
(f) The time evolution of order parameters $P$ (blue) and $M$ (red) in intermittency patterns 
between vortex and polarized school.
The parameters are $\kappa_v=1.0$, $\lambda_v=2.4$, $\mu_\theta=1.4$, $\gamma=0.6$,
and the others are the same as (a)-(e).
The time window is $t\in[80000,100000]$.
See Supplemental material Movies S6 for the dynamics.
(g) The probability distribution $\Phi(P,M)$ of intermittency
which is averaged at $t\in[300,100000]$ in a simulation.
The parameters are the same as those in (f).
}
\label{Fig2}
\end{figure*}

First, our model shows the five steady collective patterns with $N=100$ as shown in Fig.~\ref{Fig2}(a)-(e),
and they are classified as follows.
Swarm: randomly oriented agents.
Shell: rotation occurs at the center of the cluster, 
while agents enter and leave in the radial direction at the periphery.
Vortex: agents rotating around a common axis.
Polarized school: agents aligned and moving coherently in the same direction.
There are two types of polarized schools
which extend longitudinally and extend transversely with respect to the direction of motion. 
We denote them by polarized school (L) and (T), respectively. 
Unless otherwise specified, (L) is implied.

The size of these clusters can reach several dozen.
~\ref{FigE1} shows the typical time evolution of polar order parameter $P(t)$,
rotational order parameter $M(t)$, radius of gyration $R_g(t)$,
and direction-shape correlations $q_1(t),~q_2(t)$ of a cluster.
We find that these rapidly reach a steady state and remain for a sufficiently long time.
See the Methods section ``Measured quantities'' for definitions of these and the quantities introduced below.

Furthermore, when intermediate parameter values 
between those producing the vortex and polarized school are selected, 
the cluster shows intermittency of patterns with time interval $\sim\mathcal{O}(1000)$ (Fig.~\ref{Fig2}(f)).
In fact, the probability distribution function of order parameters $\Phi(P,M)$ is bimodal 
in the intermittent case (Fig.~\ref{Fig2}(g)).
In addition, we show the steady vortex and polarized school with $N=1000$ 
using the same parameter values as those for $N=100$,
as shown in~\ref{FigE2}.

Next, we explain the relation between collective patterns and the parameters.
The parameters related to cluster cohesion are $\kappa_\theta$, $\lambda_\theta$, and $D_\theta$.
As shown in~\ref{FigE3}(b),(d),(e), small $\kappa_\theta$ and large $\lambda_\theta,~D_\theta$ cause
cohesion with small $R_g$, and then the cluster becomes a swarm.
The noise $D_\theta$ induces cohesion in the cluster
because disordered motion of agents generates a large optic flow.
Moreover, the order parameters do not change appreciably 
with these parameters (see~\ref{FigE3}(c)), 
and we adopt $\kappa_\theta=1.0$, $D_\theta=0.005$ and mainly use $\lambda_\theta=3.5$
at which a stable cohesive state begins to emerge.

The parameter $\kappa_v$ controls collisions and 
overlaps between agents.
When $\kappa_v$ increases,
the probability of overlap $p_o$ and the mean speed $U$ of agents decrease,
and the shell emerges (see~\ref{FigE4}(b)-(d)).
In other words, 
as the speed decreases, the radius of curvature of the turning of an agent becomes smaller 
for the angular velocity given by Eq.~(\ref{eqtheta}),
and then we qualitatively find that 
localized turning with little movement in the direction of motion 
facilitates the formation of rotational patterns.
Indeed, comparing $\kappa_v=0$ and $\kappa_v=1.0$, 
rotation is more strongly induced in the $\kappa_v=1.0$ case
(see~\ref{FigE3}(f),~\ref{FigE4}(f)).
As shown in \ref{FigE3}(e),
the characteristic of the shell pattern is that 
agents tend to stagnate at the periphery of the cluster, 
where density increases and speed decreases.
Note that the noise parameter $D_v$ has little effect on the patterns.
We adopt $\kappa_v=1.0,~D_v=0.001$
where the speed does not decrease drastically, 
while $p_o$ decreases to approximately $10^{-3}$.

The remaining parameters are $\lambda_v$, $\mu_v$, $\mu_\theta$, $\gamma$.
As shown in Fig.~\ref{Fig3}(a),
when $\lambda_v$ increases, the school emerges and then polar order parameter $P$ increases.
In other words, large $\lambda_v$ facilitates the alignment between agents.
Furthermore, as $\lambda_v$ increases, 
the shape changes from the polarized school (L) ($q_2\approx1$)
to the polarized school (T) ($q_1\approx1$).
This can be interpreted as follows: 
increased sensitivity to front-back positional adjustments causes the cluster 
to spread more in the lateral direction.
At $\lambda_v=1.0$, 
when $\mu_\theta,\gamma$ increase,
the vortex emerges (see Fig.~\ref{Fig3}(b)).
Moreover, we find that,
in the ordered cluster, the mean speed $U$ reaches 1
and the probability of overlap $p_o$ reaches $\mathcal{O}(10^{-5})$ (see Fig.~\ref{Fig3}(a)-(b) and ~\ref{FigE5}(a)).
This indicates that the agents move at a steady speed $v_0$ 
by minimizing the effects of positional variation.
In addition, the intermittency pattern emerges at $\lambda_v=2.0$
where $P\approx M$ and the duration of orders $TP\approx TM$,
and the school is maintained at $\lambda_v=3.0$ when $\mu_\theta,\gamma$ are changed
(see~\ref{FigE5}(b)-(c)).
For example, as $\mu_\theta$ increases, 
the pattern changes from a polarized school to an intermittency pattern and then to a vortex
at $\lambda_v=2.4$ (see Fig.~\ref{Fig3}(d)).
Note that, as shown in Fig.~\ref{Fig3}(c),
the magnitude of $\mu_v$ has little effect on the patterns
because the orientations of agents are locally aligned,
and we set $\mu_v=0$, unless specifically mentioned.

\section*{Scale-free correlation and topological distance}

To reveal the properties of the topological interaction, 
which depends on the relative positions between agents rather than the distance,
we measure the correlation function of velocity fluctuations $C(r)$
for the cluster with relatively large polar order.
See the Methods section ``correlation function" for the definition.
As shown in Fig.~\ref{Fig4}(a),
$C(r)$ exhibits damped oscillations at short range,
and then reaches zero at the correlation length $\xi$.
We find that clusters with large polar order have 
large $\xi$,
and that $\xi$ depends linearly on the cluster gyration radius $R_g$ for $N\in[20,500]$
(see Fig.~\ref{Fig4}(b)-(c)):
the clusters exhibit scale-free correlations.
On the other hand,
the probability distributions of the minimum distance between agents $P(d)$
do not change significantly when $N$ is changed (see Fig.~\ref{Fig4}(d)),
indicating that the local number density does not change markedly.

Next, we consider the topological range:
how many neighbors are actually visible from an agent.
We measure the occupancy angle $\Theta_{tot}$ as the angular
region of the images in the field of view (see Fig.~\ref{Fig4}(e)).
$\Theta_{tot}$ increases when $N$ increases (see Fig.~\ref{Fig4}(f)),
and the images occupy about 75-85\% of the field of view.
Furthermore, we define the relative occupancy angle $\Theta_n$ 
as the angular region occupied by the image of the $n$-th nearest neighbor ($n$NN),
where $n$ is topological distance.
As shown in Fig.~\ref{Fig4}(g),
for all collective patterns with $N=100$ and 200, 
$\Theta_n$ decreases rapidly, 
and $\Theta_n/\Theta_1$ reaches approximately 0.2 at $n=10$.
$\Theta_n/\Theta_1$ is smaller for $N=100$ than for $N=200$,
because the distance to $n$NN becomes larger for smaller $N$ due to finite-size effects.
Here, we introduce $\Theta_n$ when the agents are randomly distributed as
\begin{eqnarray}
\label{Thetarand}
&&\Theta_{\mathrm{rand}}(n;\chi_1,\chi_2)=C_\Theta\tan[-1](\frac{\chi_2}{\sqrt{n}})\\\nonumber
&&\times\exp[-\chi_1\leri{(n+\chi_2^2)\tan[-1](\frac{\chi_2}{\sqrt{n}})+\chi_2\sqrt{n}}],
\end{eqnarray}
where the normalization constant $C_\Theta$ 
is chosen such that
$\Theta_{\mathrm{rand}}(1;\chi_1,\chi_2)=1$
(See the Methods section ``Derivation of $\Theta_{\mathrm{rand}}(n;\chi_1,\chi_2)$").
In the case of the swarm with $N=200$, 
$\Theta_n/\Theta_1$ is consistent with $\Theta_{\mathrm{rand}}$ over the entire range of $n$.
On the other hand, 
$\Theta_n/\Theta_1$ is larger than $\Theta_{\mathrm{rand}}$ at $n\lesssim10$
for the vortex and school.
This indicates that
the visual inputs in the swarm are equivalent to random background noise,
but ordered states suppress visual occlusion at $n\lesssim10$, 
thereby allowing agents to retain distinct visual signals from neighbors.

\section*{Emergence of 3A forces}

Finally, we show the relation between the force and the parameters. 
In a two-agent system,
we measure the turning acceleration $a(t)=v\dv*{\theta}{t}$
whose direction is perpendicular to the direction of motion of an agent.
Note that the mass is set to unity, so acceleration equals force.
We set $a(r,\phi,\psi)$ as a function of distance $r$, azimuth $\phi$, and relative heading $\psi$
(see Fig.~\ref{Fig5}(a)),
and the mirror symmetry requires $a(r,-\phi,-\psi)=-a(r,\phi,\psi)$.
We assume a separable form for $a(r,\phi,\psi)$~\cite{Calovi2018},
and thus $a(r,\phi,\psi)$ is decomposed into a repulsion-attraction term (indexed 1) and an alignment term (indexed 2) as follows:
\begin{equation}
\label{funcacc}
a(r,\phi,\psi)=A_1(r)O_1(\phi)E_1(\psi)+A_2(r)E_2(\phi)O_2(\psi),
\end{equation}
where $O_1(\phi),~O_2(\psi)$ are odd functions and $E_1(\psi),~E_2(\phi)$ are even functions.
We obtain functional forms by optimizing the measured acceleration.
(See the Methods section ``Optimization of acceleration").

First, when $\lambda_\theta$ increases,
the number of samples of simulations $n(r)$ cumulated over angle region rises at a shorter distance,
the attraction part ($A_1(r)>0$) increases and 
the equilibrium distance $r_e$, defined by $A_1(r_e)=0$, shifts toward shorter distances
(see~\ref{Fig5}(b)):
it corresponds to the large cohesive force with large $\lambda_\theta$
and facilitates the swarm.

Next, we change $\lambda_v$ and $\mu_\theta$.
As shown in Fig.~\ref{Fig5},
in the case of the parameters of vortex and school,
there is little change in the repulsion-attraction term $A_1(r)O_1(\phi)E_1(\psi)$.
In both cases, $A_1(r)$ shows the equilibrium distance $r_e\sim2$, 
a peak at $r\sim10$, and sufficiently decreases by the distance to $r\sim25$.
For the case of school,
$n(r)$ has a peak around the equilibrium distance,
but for the case of vortex, 
$n(r)$ has a dispersed peak at position farther from the equilibrium distance.
In the case of school,
two agents travel parallel to each other and then
their motion stabilizes near the static equilibrium distance $r_e$.
In contrast, in the case of vortex,
mutual angular variation induces rotational dynamics,
and two agents oscillate back and forth away from $r_e$.
(See Supplemental material Movies S9-S10 for the dynamics.)
In addition, the magnitude of $O_1(\phi)$ has a peak at $\phi\sim\pm100^\circ$,
but $E_1(\psi)$ takes 1 almost independently of $\psi$.

Furthermore, we show the effect of $\lambda_v$ and $\mu_\theta$ 
on the alignment term $A_2(r)E_2(\phi)O_2(\psi)$
(see Fig.~\ref{Fig5}(d)).
$A_2(r)$ has the non-zero values within the equilibrium distance $r_e$,
and its peak increases when $\lambda_v$ increases and $\mu_\theta$ decreases.
In other words, rather than rotating to minimize the optic flow,
accelerating toward a strong optic flow and following the neighbor 
facilitates alignment.
$E_2(\phi)$ shows that the neighbor in the lateral direction is more strongly referenced 
when $\lambda_v$ increases, 
whereas the forward neighbor is weighted more strongly 
when $\mu_\theta$ increases.
$O_2(\psi)$ is relatively insensitive to $\lambda_v$, 
but increasing $\mu_\theta$ broadens the width of $O_2(\psi)$ around $\psi\sim\pm90^\circ$.
Note that the effect of $\mu_v$ is negligible
because overtaking and passing each other occur rarely after relaxation in a simulation
(see~\ref{FigE6}).

\section*{Discussion}
In summary, we present a model based on biological visual detection
that does not prescribe a static equilibrium distance.
We introduce simple interactions derived from system symmetries and low-order function expansions,
and add a feedforward inhibitory mechanism for the input.
In fact, without the inhibition, the cluster becomes strongly cohesive and loses order
(see Supplementary Section II for the details),
and the previous visual model shows the saturation of the visual field by overlap of agents~\cite{Bastien2020}.
This suggests that selectivity for stimulus intensity~\cite{Preuss2014} is also involved in cluster formation.

Our model shows the ordered patterns:
the vortex and the polarized school (L,~T).
Fish mainly form the polarized school (L)~\cite{Pitcher1979},
and show the intermittency patterns between polarized school (L) and vortex~\cite{Tunstrom2013}:
rotation begins when agents at the front edge of an aligned cluster move around to the lateral rear of the cluster
(see Supplemental material Movies S6). 
On the other hand, birds show both the polarized school (L) and (T) 
by turning with the relative positional relationships among individuals maintained~\cite{Ballerini2008b,Hemelrijk2011}.
In addition, the shell is a previously unobserved pattern, 
in which agents are adiabatically drawn toward the center of the cluster
due to reduced optic flow resulting from the low agent speed under strong repulsion.

Our model exhibits scale-free correlations ($\xi\approx R_g$),
which are a common property of models with topological interactions~\cite{Hemelrijk2011,Hang2026}.
The ratio of the visual field which is not occupied by the other agents is 15-25\% for $N=100$-200, 
which is consistent with the ratio 10-40\% for 151 golden shiners~\cite{Davidson2021}.
Furthermore, in the ordered cluster, neighbors at a topological distance of $n\lesssim10$
are less occluded than in the swarm.
This is consistent with observed topological interaction ranges of 1~\cite{Herbert2011}, 
3 or 6-8~\cite{Gautrais2012} in fish and 6 in birds~\cite{Ballerini2008}.
Notably, the scale-free correlations emerge despite the interaction being restricted to 
a small topological distance ($n\lesssim10$),
which is not a prescribed rule 
(e.g. interaction capacity~\cite{Ito2022,Hemelrijk2011} and Voronoi tessellation~\cite{Gautrais2012,Calovi2014}) 
but spontaneously emerges from visual occlusion.

We decompose the turning acceleration into the attraction term and the alignment term.
In the attraction term,
$A_1(r)$ has a significant value up to a distance $r\sim25$,
which corresponds to the distance $r\sim16$-39 associated with the visual resolution of golden shiners~\cite{Pita2015}.
Depending on the value of $\lambda_\theta$, the equilibrium distance is statistically obtained as $r_e\sim2$.
However, unlike models with a pre-defined isotropic potential,
interactions are determined dynamically,
and then equilibrium does not necessarily guarantee dynamic stability as seen in the parameter case for the vortex.
The peak angle of $\abs{O_1(\phi)}$ ($\sim\pm100^\circ$) is consistent with 
the results of the fish experiments~\cite{Calovi2018},
but $E_1(\psi)$ shows poor agreement with experiment:
$\abs{J}$ is not well suited to reflecting the relative orientation of a neighbor 
because of the cylindrical shape of agents.

In the alignment term,
$A_2(r)$ increases not with the parameters $\mu_v,~\mu_\theta$ for directional synchronization to the optic flow, 
which reflects velocity matching with the neighbor~\cite{Castro2024}, 
but with the parameter $\lambda_v$ for approaching regions of strong optic flow (see Fig.~\ref{Fig5}(e)).
This mechanism is similar to those in which an aligned cluster emerges through potential-driven attraction 
without explicitly introducing alignment interactions~\cite{DOrsogna2006}.
However, their underlying dynamic mechanisms differ fundamentally. 
Potential-driven forces directly reorient agents toward a static equilibrium where forces and torques vanish. 
In contrast, our model maintains alignment through a sensorimotor feedback loop: 
alignment suppresses optic flow, 
but residual repulsion causes minor displacements that regenerate optic flow, 
thereby continuously reactivating the $\lambda_v$-mediated approach response and stabilizing alignment.
Note that, unlike observations in fish~\cite{Calovi2018}, $A_2(r)$ increases within the equilibrium distance of $A_1(r)$.
This discrepancy may be resolved by incorporating predictive mechanisms, 
such as temporal memory of optic flow~\cite{Bahl2020,Dragomir2020} 
and social experience dependent learning of alignment~\cite{Zada2024}:
earlier prediction could enhance alignment interactions at larger distances around $r_e$.
As indicated by the changes in $E_2(\phi)$ and $O_2(\psi)$, 
the vortex is achieved through $\mu_\theta$ by following the optic flow of forward neighbors 
and referencing a broader range of relative heading angles -- 
a trend consistent with experiment~\cite{Calovi2018}.

In conclusion, our study highlights the importance of sensory processing 
based on optic flow in collective motion. 
We demonstrate that the phenomenological 3A rules emerge as statistical averages of microscopic interactions, 
and reveal their underlying origin in dynamic and flexible interactions driven by optic flow. 
The minimal nature of our framework provides a tractable baseline.
We anticipate that this work will serve as a stepping stone for future models 
that build higher-level processes, 
such as sensory selectivity~\cite{Sridhar2021} and predictive inference~\cite{Heins2024}
based on elementary visual cues.
Therefore, we welcome further work in these areas.

\section*{Code and Data Availability}
The simulation code and data supporting the findings of this study are available at
\url{https://doi.org/10.5281/zenodo.23786429}

\begin{acknowledgments}
We acknowledge financial support by JSPS KAKENHI Grant No. 26KJ0017 to S.I.
\end{acknowledgments}



\begin{figure*}[!t]
\centering
\includegraphics[width=\linewidth]{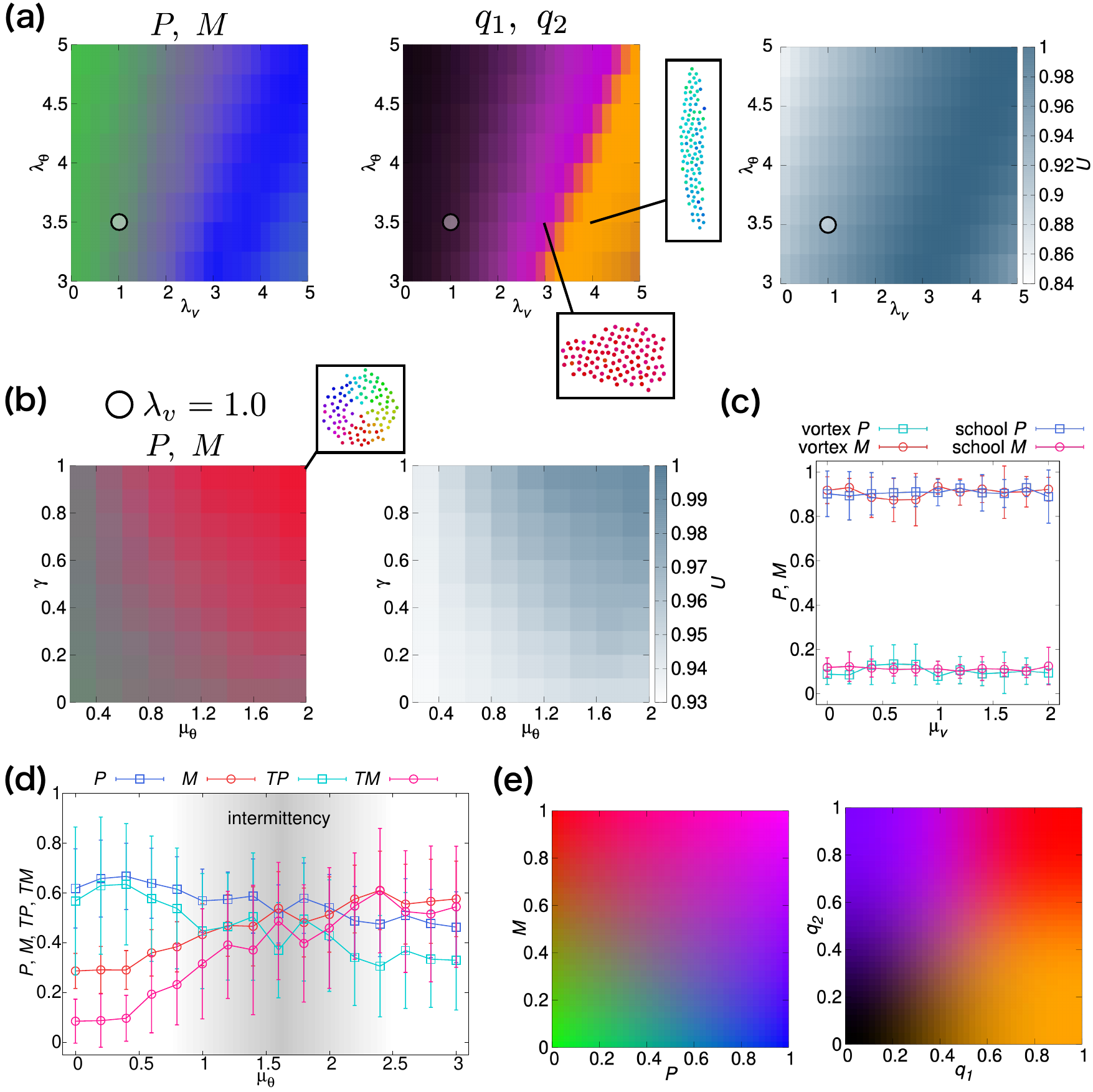}
\caption{
Phase diagram of collective patterns with $N=100$.
$\kappa_\theta=1.0$, $\kappa_v=1.0$, $D_v=0.001$, and $D_\theta=0.005$ are fixed.
(a) Dependence of patterns on $\lambda_v$ and $\lambda_\theta$
with $\mu_v=\mu_\theta=\gamma=0$:
the order parameters $P,M$ (left), the direction-shape correlations $q_1,q_2$ (center), and the mean speed $U$ (right).
The polarized school (L) emerges at $\lambda_v=3.0$ and the polarized school (T) emerges at $\lambda_v=4.0$
with $\lambda_\theta=3.5$ (see Fig.~\ref{Fig2}(d),(e), respectively).
The circle point indicates $\lambda_v=1.0$, $\lambda_\theta=3.5$ 
which is the starting parameter point of the diagram in (b).
(b) Dependence of patterns on $\mu_\theta$ and $\gamma$ with $\mu_v=0$:
the order parameters $P,M$ (left) and the mean speed $U$ (right).
The vortex emerges at $\mu_\theta=2.0$ and $\gamma=1.0$ (see Fig.~\ref{Fig2}(c)).
(c) Dependence of the order parameters $P,M$ on $\mu_v$.
The cyan line is $P$ and the red line is $M$ of the vortex,
and the blue line is $P$ and the pink line is $M$ of the polarized school.
(d) The order parameters $P$ (blue), $M$ (red) and the duration of order $TP$ (cyan), $TM$ (pink) 
of the intermittency pattern
between the vortex and the polarized school
with $\lambda_v=2.4$, $\lambda_\theta=3.5$, $\mu_v=0$, $\gamma=0.6$.
The gray-shaded area in $\mu_\theta\sim1$-2.5 corresponds to the intermittency pattern 
($P\approx M$ and $TP\approx TM$).
The case of $\mu_\theta=1.4$ corresponds to the parameters in Fig.~\ref{Fig2}(f)-(g).
In (c) and (d), the error bars represent the standard deviation over 25 simulations 
for the time-averaged values at $t\in[300,1000]$.
(e) The color represents the value of order parameters $P,M$ (left) 
and the value of direction-shape correlations $q_1,q_2$ (right).
}
\label{Fig3}
\end{figure*}

\begin{figure*}[!t]
\centering
\includegraphics[width=\linewidth]{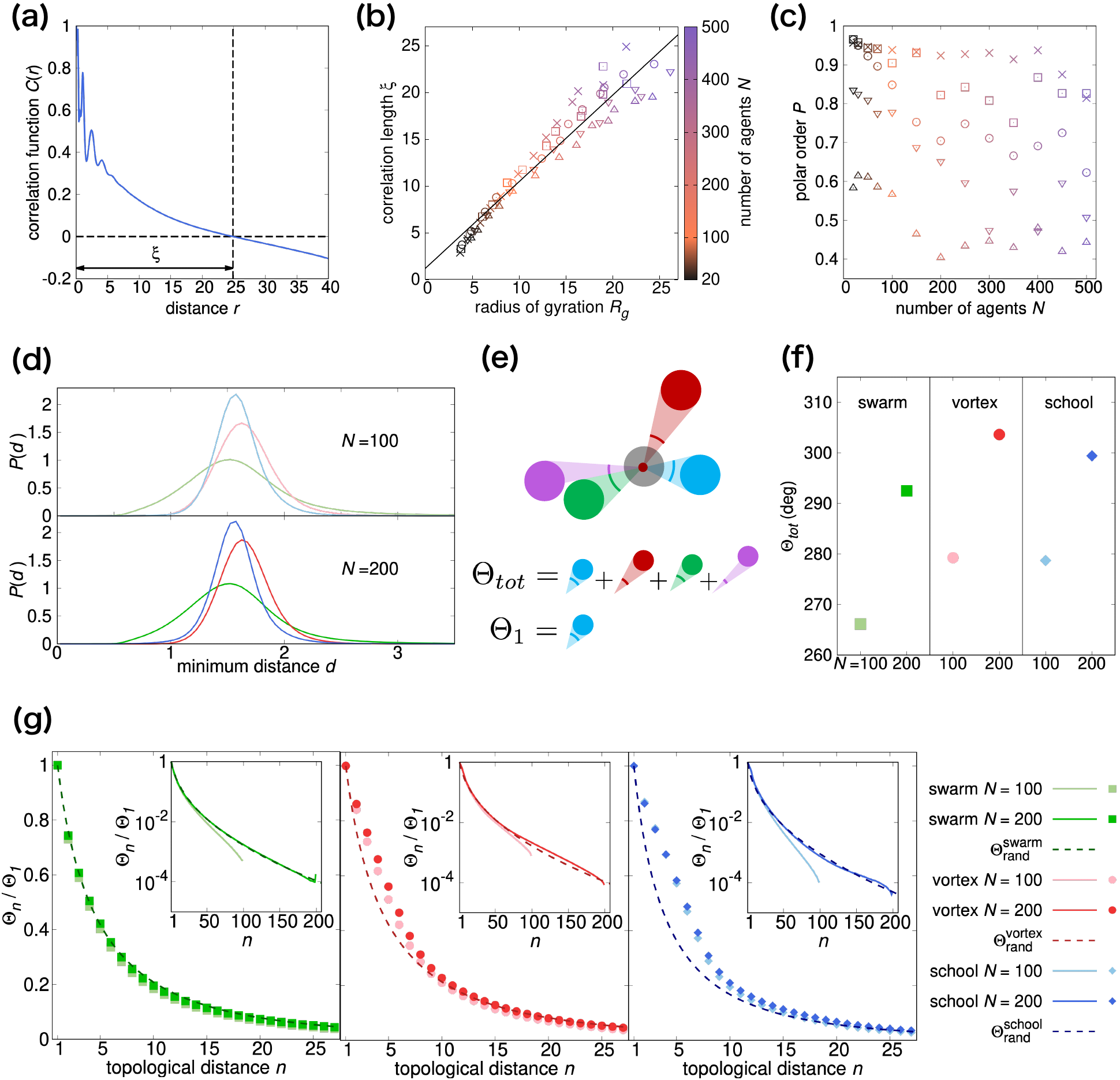}
\caption{
Scaling law and topological interaction.
The time averaging is performed at $t\in[300,20000]$ in a simulation for each parameter to obtain (a)-(d) and (f)-(g).
(a) The correlation function of velocity fluctuations  $C(r)$ as a function of the distance from each agent.
The parameters are $N=500$, $\lambda_v=3.0$, $\mu_\theta=\gamma=0$ which are the same as
those of the polarized school with $N=100$ (Fig.~\ref{Fig3}(d)).
The black arrow shows the correlation length $\xi$.
(b) The relation between the correlation length $\xi$ and the radius of gyration $R_g$
with $N=20$-500.
Cross: $\lambda_v=3.0$, $\mu_\theta=\gamma=0$.
Square: $\lambda_v=3.0$, $\mu_\theta=1.0$, $\gamma=0.6$.
Circle: $\lambda_v=3.0$, $\mu_\theta=2.0$, $\gamma=0.6$.
Upward triangle: $\lambda_v=2.0$, $\mu_\theta=\gamma=0$.
Downward triangle: $\lambda_v=2.5$, $\mu_\theta=\gamma=0$.
These points are fitted by the linear function $\xi=\hat{\xi} R_g+\xi_0$ (black line):
$\hat{\xi}=0.925\pm0.034$, $\xi_0=1.192\pm0.491$.
(c) Dependence of the polar order parameter $P$ on the number of agents $N$.
The meaning of the point shapes is the same as in (b).
(d) The probability distribution function $P(d)$ as a function of the minimum distance $d$
with $N=100$ (upper) and $N=200$ (lower).
Swarm (green): $\lambda_v=1.0$, $\mu_\theta=\gamma=0$.
Vortex (red): $\lambda_v=1.0$, $\mu_\theta=2.0$, $\gamma=1.0$.
School (blue): $\lambda_v=3.0$, $\mu_\theta=\gamma=0$.
See also the list of legends for (d), (f)-(g) at the bottom right of this figure.
(e) Schematic figure of $\Theta_{tot}$ and $\Theta_1$.
(f) $\Theta_{tot}$ in swarm (left), vortex (center), and school (right) 
with $N=100$ and 200.
(g) The relative occupancy angle $\Theta_n$ normalized by $\Theta_1$
in swarm (left), vortex (center), and school (right).
The insets show the semi-log plots $\Theta_n/\Theta_1$ in each figure.
The dashed lines are the fitting curves of normalized $\Theta_{\mathrm{rand}}(n;\chi_1,\chi_2)$ for $N=200$:
$\chi_1=0.132\pm0.001$, $\chi_2=2.22\pm0.03$ (swarm),
$\chi_1=0.129\pm0.002$, $\chi_2=2.22\pm0.06$ (vortex),
$\chi_1=0.170\pm0.004$, $\chi_2=1.85\pm0.07$ (school).
}
\label{Fig4}
\end{figure*}

\begin{figure*}[!t]
\centering
\includegraphics[width=\linewidth]{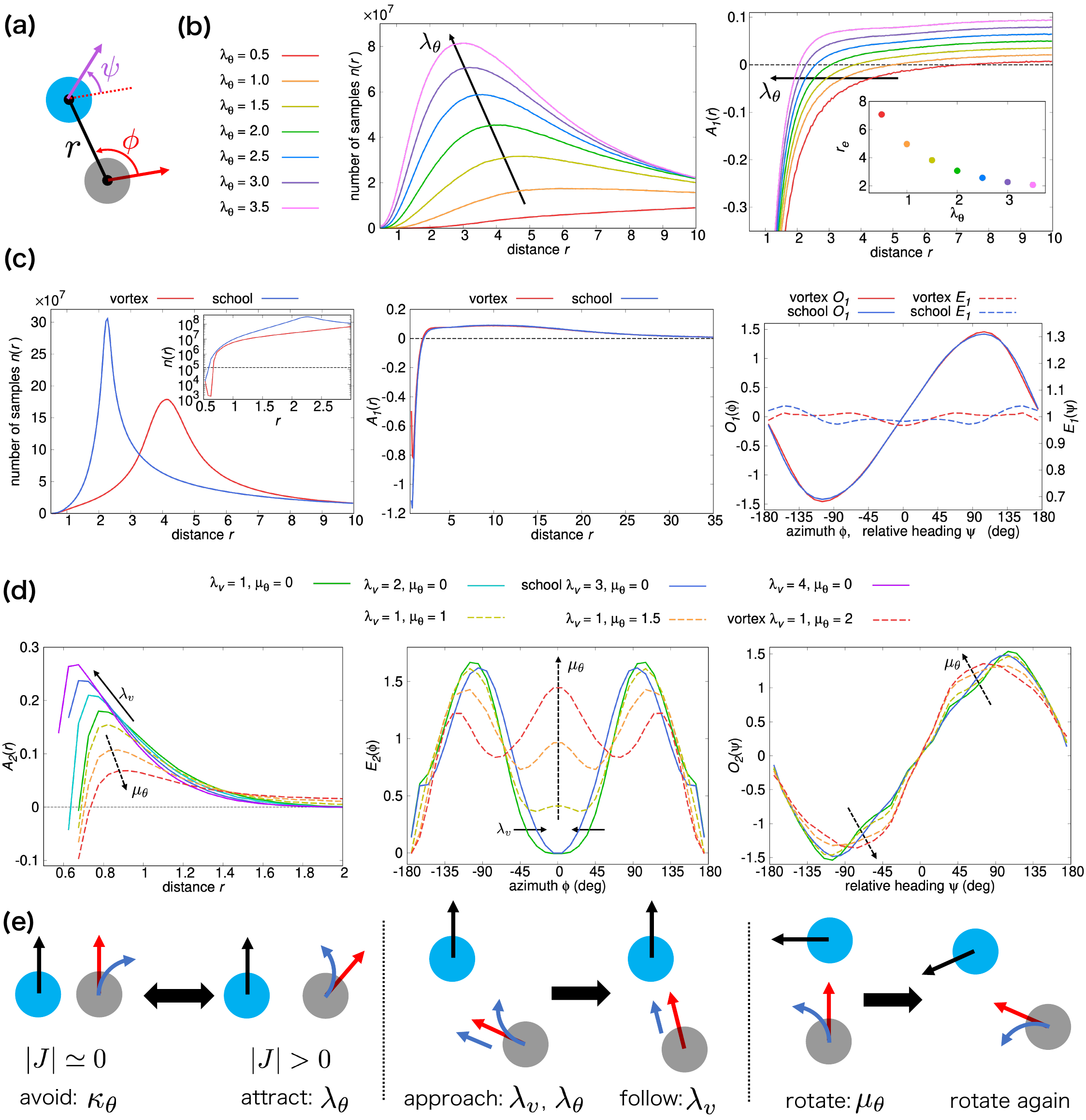}
\caption{
Decomposed turning acceleration.
(a) Schematic figure of distance $r$, azimuth $\phi$, and relative heading $\psi$.
The arrows represent the direction of motion of each agent.
(b) Dependence of the number of samples $n(r)$ (left) and $A_1(r)$ (right) on $\lambda_\theta\in[0.5,3.5]$
with $\kappa_\theta=1.0$, $\kappa_v=\lambda_v=\mu_\theta=\gamma=0$.
$n(r)$ corresponds to the summation $\sum_{lm}n_{klm}$ over angular bins
(see the Methods section ``Optimization of acceleration").
In the right panel, the inset shows dependence of the equilibrium distance $r_e$ on $\lambda_\theta$.
The colors of the points are the same as those of the lines in the legend.
(c) The repulsion-attraction term with the parameters for 
the vortex ($\lambda_v=1.0$, $\mu_\theta=2.0$, $\gamma=1.0$; red)
and the school ($\lambda_v=3.0$, $\mu_\theta=\gamma=0$; blue)
which are the same as those in Fig.~\ref{Fig3}(c) and (d), respectively.
$n(r)$ is left, $A_1(r)$ is center, and $O_1(\phi)$ (solid line), $E_1(\psi)$ (dashed line) is right.
In the left panel, the inset shows the semi-log plot of $n(r)$ at $r\in[0.5,3]$
and the black dashed line represents the threshold $n=129600$.
$n(r)$ is below the threshold only at distance very close to 0.5
due to the effect of avoidance
(see also the Methods section ``Optimization of acceleration").
(d) The alignment term with changes in $\lambda_v$ and $\mu_\theta$ ($\gamma=1.0$ is fixed).
The solid lines are $\lambda_v=1.0$ (green), $\lambda_v=2.0$ (cyan), $\lambda_v=3.0$ (blue; school),
and $\lambda_v=4.0$ (purple) with $\mu_\theta=0$.
The dashed lines are $\mu_\theta=1.0$ (yellow), $\mu_\theta=1.5$ (orange), 
and $\mu_\theta=2.0$ (red; vortex) with $\lambda_v=1.0$.
$A_2(r)$ is left, $E_2(\phi)$ is center, and $O_2(\psi)$ is right.
(e) Schematic illustration of the elements of interaction.
Left: the competition between the effect of $\kappa_\theta$ and $\lambda_\theta$ induces 
oscillation around the equilibrium distance as $\abs{J}$ varies.
Center: the effects of $\lambda_v$ and $\lambda_\theta$ cause the agent to approach its neighbor, 
and then the effect of $\lambda_\theta$ on the forward neighbor becomes weaker, 
and $\lambda_v$ induces following, resulting in an apparent alignment.
Right: the effect of $\mu_\theta$ induces repeated rotations 
in response to the flow of the forward neighbor, thereby promoting the vortex.
}
\label{Fig5}
\end{figure*}

\clearpage
\setcounter{figure}{0}
\renewcommand{\thefigure}{Extended Data Fig.~\arabic{figure}}
\renewcommand{\theHfigure}{ExtendedDataFig.\arabic{figure}}
\makeatletter
\def\fnum@figure{\thefigure}
\makeatother
\section*{Methods}

\subsection*{Derivation of $J(\phi,t)$}

The HRC was initially developed as an empirical model for an insect \textit{Chlorophanus},
and was later applied to describe the activity of adjacent receptive fields in flies~\cite{Egelhaaf1987}.
In this model, a signal $f$ from a moving image is received 
by two receptive fields located at angular positions $\phi$ and $\phi+\varphi$ ($\varphi>0$) in the visual field. 
The signal from one receptive field is delayed by a time $\tau$ and 
multiplied by the instantaneous signal from the other receptive field.
Two receptive fields are paired, 
and each pair detects positive or negative optic flow according to its preferred direction.
The optic flow in the positive direction is computed as $f(\phi,t-\tau)f(\phi+\varphi,t)$, 
and the optic flow in the negative direction is computed as $f(\phi,t)f(\phi+\varphi,t-\tau)$,
and then the total detector signal is represented as 
\begin{equation}
\label{outf}
f(\phi,t-\tau)f(\phi+\varphi,t)-f(\phi,t)f(\phi+\varphi,t-\tau).
\end{equation}

We expand Eq.~(\ref{outf}) with $\varphi$ and $\tau$ as small parameters~\cite{Egelhaaf1987}, 
and choose $f(\phi,t)$ as the vertical angular diameter $\delta(\phi,t)$
detected by vertically arranged HRC at $\phi$ in the visual system.
Thus, we obtain
\begin{eqnarray}
\label{}
&&\delta(\phi,t-\tau)\delta(\phi+\varphi,t)-\delta(\phi,t)\delta(\phi+\varphi,t-\tau)\\\nonumber
&\simeq&\varphi\tau\leri{\delta\pdv{\delta}{\phi}{t}-\pdv{\delta}{\phi}\pdv{\delta}{t}},
\end{eqnarray}
and then we define $J(\phi,t)$ as Eq.~(\ref{eqJ}).
The constant $\varphi\tau$ is absorbed into the parameters multiplying $J(\phi,t)$ in our model.

This formulation can also be applied to the mechanism of optic flow detection based on DS-RGC in vertebrates.
As shown in~\ref{FigE7}, 
a DS-RGC receives spatially asymmetrical inhibitory connections from starburst amacrine cells (SACs). 
This asymmetrical wiring forms the preferred direction~\cite{Yonehara2011}.
On the dendrites of the DS-RGC, 
the delayed signal from the SAC is multiplied 
with the instantaneous signal from bipolar cells (upstream excitatory cells).
Consequently, combining the responses of a pair of DS-RGCs with opposite preferred directions 
yields an output signal equivalent to Eq.~(\ref{outf}).

\subsection*{Determination of the interaction functions}
We determine the functional form of the interaction functions 
$W_v,~W_\theta,~Z_v,~Z_\theta$ in Eqs.~(\ref{eqv},~\ref{eqtheta}).
We consider a configuration by mirror symmetry of images:
$\delta(\phi)\to\delta(-\phi)$ and $J(\phi)\to-J(-\phi)$.
In this case, $\dv*{v}{t}\to\dv*{v}{t}$ 
and $\dv*{\theta}{t}\to-\dv*{\theta}{t}$ in Eqs.~(\ref{eqv},~\ref{eqtheta}) are necessary.
Therefore, due to
\begin{eqnarray}
\label{}
&&\int_{-\pi}^\pi\dd\phi\{W_v(\delta(-\phi),\phi)+Z_v(-J(-\phi),\phi)\}\\\nonumber
&=&\int_{-\pi}^\pi\dd\phi\{W_v(\delta(\phi),-\phi)+Z_v(-J(\phi),-\phi)\}
\end{eqnarray}
and similarly for $W_\theta$ and $Z_\theta$,
the following conditions are required:
\begin{equation}
\label{condW}
W_v(\delta,\phi)=W_v(\delta,-\phi),~W_\theta(\delta,\phi)=-W_\theta(\delta,-\phi)
\end{equation}
\begin{equation}
\label{condZ}
Z_v(J,\phi)=Z_v(-J,-\phi),~Z_\theta(J,\phi)=-Z_\theta(-J,-\phi).
\end{equation}
We assume a separable form for the functions,
and then the parity of the function of each variable is determined 
by these conditions~(\ref{condW},~\ref{condZ}):
\begin{equation}
\label{}
W_v(\delta,\phi)=K_v(\delta)\Gamma_{K,e}(\phi),
\end{equation}
\begin{equation}
\label{}
W_\theta(\delta,\phi)=K_\theta(\delta)\Gamma_{K,o}(\phi),
\end{equation}
\begin{equation}
\label{}
Z_v(J,\phi)=L_v(\abs{J})\Gamma_{L,e}(\phi)+\sgn{J}M_v(\abs{J})\Gamma_{M,o}(\phi),
\end{equation}
\begin{equation}
\label{}
Z_\theta(J,\phi)=L_\theta(\abs{J})\Gamma_{L,o}(\phi)+\sgn{J}M_\theta(\abs{J})\Gamma_{M,e}(\phi),
\end{equation}
where $\sgn{x}$ is the sign function ($\sgn{x\geq0}=1,~\sgn{x<0}=-1$),
and the indices $e$ and $o$ for 
each $\Gamma(\phi)$ function denote the even and odd functions of $\phi$, respectively.

From here, we give the explicit form of the functions.
We choose the linear function of image size for avoidance:
\begin{equation}
\label{}
K_v(\delta)=-\kappa_v\delta,~K_\theta(\delta)=-\kappa_\theta\delta.
\end{equation}
Indeed, the probability of occurrence of avoidance is proportional to 
the retinal image size $\int_{-\pi}^\pi\dd\phi\delta(\phi)$~\cite{Harpaz2021}.

Next, we show the $J$-dependent functions.
We derive an inhibitory system for detecting the weak stimuli
by using the well-studied mechanism of SINs in the fish tectum.
The SINs respond to temporally varying stimuli from moving images~\cite{Yin2019},
and mediate feedforward inhibition in downstream neural circuits~\cite{Preuss2014}.
We define the action potential of downstream neural circuits $V_{neu}$ and
the amount of inhibitory substances $N_{inh}$ from SINs,
and assume the relations $V_{neu}\propto\abs{J},~N_{inh}\propto\abs{J}$, and $\dd V_{neu}\propto-V_{neu}\dd N_{inh}$.
Thus, we obtain $V_{neu}\propto\abs{J}\exp(-\abs{J}/J_0)$.
Indeed, the SINs exhibit the exponential inhibition $\exp{-\mathrm{const.}\times(\mathrm{firing~rate})}$~\cite{Dunn2016}.
Therefore, we define
\begin{equation}
\label{}
L_v(\abs{J})=\lambda_v\abs{J}e^{-\frac{\abs{J}}{J_0}},~L_\theta(\abs{J})=\lambda_\theta\abs{J}e^{-\frac{\abs{J}}{J_0}},
\end{equation}
\begin{equation}
\label{}
M_v(\abs{J})=-\mu_v\abs{J}e^{-\frac{\abs{J}}{J_0}},~M_\theta(\abs{J})=\mu_\theta\abs{J}e^{-\frac{\abs{J}}{J_0}}.
\end{equation}

Finally, we expand the $\Gamma(\phi)$ functions with the lower-order terms:
\begin{equation}
\label{}
\Gamma_{K,o}(\phi)=\Gamma_{L,o}(\phi)=\Gamma_{M,o}(\phi)=\sin\phi,
\end{equation}
\begin{equation}
\label{}
\Gamma_{K,e}(\phi)=\Gamma_{L,e}(\phi)=\cos\phi,~\Gamma_{M,e}(\phi)=1+\gamma\cos\phi,
\end{equation}
where, when $\delta(\phi)$ and $J(\phi)$ are constants, 
only the interaction that rotates the agent according to the optic flow arises and other interactions do not emerge due to symmetry,
and then $\Gamma_{M,e}(\phi)$ is expanded from a constant term.
From above, we obtain Eqs.~(\ref{funcWv},~\ref{funcWt},~\ref{funcZv},~\ref{funcZt}).

\subsection*{Numerical methods}

\subsubsection*{initial condition and time steps}
Unless otherwise noted for the initial condition, 
we randomly positioned $N$ agents within a circle of radius $10\times\sqrt{N/100}$
because we set the density to remain constant based on 100 agents.
The agents have the same initial speed $v_i=v_0$ 
with the random moving directions $\theta_i$ uniformly distributed in $[-\pi,\pi]$.

We numerically integrated Eqs.~(\ref{eqv},~\ref{eqtheta}) by the Euler method
with the time step $\Delta t=0.005$, and the standard Wiener processes are replaced by $\xi_w\sqrt{\Delta t}$
with a random number $\xi_w$ generated by the standard normal distribution.

\subsubsection*{visual field and angular diameter}
The angle of visual field $\phi\in[-\pi,\pi]$ is divided into bins with the angular width $\Delta\phi=0.5^\circ$.
The angular diameter $\delta_i(\phi)$ is calculated as follows.
We introduce a unit vector $\bm{\varepsilon}_i(\phi)=\{\cos(\theta_i+\phi),\sin(\theta_i+\phi)\}$
and the relative position vector $\bm{r}_{ij}=\bm{r}_j-\bm{r}_i$ for neighbor agent $j$,
and then we obtain the distance $\mathfrak{r}_{ij}(\phi)$ to the intersection with neighbor
when $\bm{\varepsilon}_i(\phi)$ is extended from the eye:
\begin{equation}
\label{}
\mathfrak{r}_{ij}(\phi)=\bm{r}_{ij}\cdot\bm{\varepsilon}_i(\phi)-\sqrt{\{\bm{r}_{ij}\cdot\bm{\varepsilon}_i(\phi)\}^2-\bm{r}_{ij}^2+\leri{\frac{l}{2}}^2}.
\end{equation}
Thus, the angular diameter is defined as
\begin{equation}
\label{}
\delta_i(\phi)=2\arctan(\frac{h}{2\tilde{\mathfrak{r}}_{i}(\phi)}),
\end{equation}
where $\tilde{\mathfrak{r}}_{i}(\phi)=\min_{j\neq i}\mathfrak{r}_{ij}(\phi)$ 
by considering the visual occlusion effects caused by the relative positions of neighbors.

Next, to calculate the derivative of angular diameter $\pdv*{\delta_i}{\phi}$,
we apply the Gaussian blur to $\delta_i(\phi)$ to prevent the nondifferentiability at the edge of the image.
Indeed, the receptive fields in the visual system cause the blurring of spatial information with a certain width.
We define the Gaussian function $G(\phi)=\exp{-\phi^2/(2\sigma^2)}$, 
and then we obtain the Gaussian-blurred angular diameter as
\begin{equation}
\label{}
\tilde{\delta}_i(\phi)=\frac{\int_{\phi-3\sigma}^{\phi+3\sigma}\dd\phi'\delta_i(\phi')G(\phi-\phi')}{\int_{\phi-3\sigma}^{\phi+3\sigma}\dd\phi'G(\phi-\phi')},
\end{equation}
where the integration range is truncated to a range of $6\sigma$ centered on $\phi$.
We set $\sigma=2^\circ$: for example, 
the full width at half maximum of the receptive field of RGC in goldfish $2\sqrt{2\ln2}\sigma$ is $\sim2$-$6^\circ$~\cite{Johnston2015},
and the retinal resolution angle in golden shiners is $\sim1$-$4^\circ$~\cite{Pita2015,Ito2024}.
We denote $\tilde{\delta}_i(\phi)$ as $\delta_i(\phi)$ elsewhere in the paper.

Finally, we consider the case where the eye enters the body of a neighbor ($r_{ij}=\abs{\bm{r}_{ij}}<l/2$).
In this case, we set the interaction functions $W_v,~W_\theta,~Z_v,~Z_\theta$ to zero,
and the time derivative of angular diameter is calculated from the time step $t+\Delta t$ 
after the overlap has been resolved at $t$.
Note that overlaps rarely occur in an ordered collective motion
(see the main text).

\subsection*{Measured quantities}
\subsubsection*{quantities for phase diagrams}
The polar order parameter is
\begin{equation}
\label{}
P(t)=\abs{\bm{P}(t)}=\frac{1}{N}\abs{\sum_{i=1}^N\bm{e}_i(t)},
\end{equation}
where $\bm{P}(t)$ is the vector polar order parameter.
The rotational order parameter is
\begin{equation}
\label{}
M(t)=\frac{1}{N}\abs{\sum_{i=1}^N\hat{\bm{c}}_i(t)\times\bm{e}_i(t)},
\end{equation}
where $\hat{\bm{c}}_i(t)=(\bm{r}_i(t)-\bm{r}_G(t))/\abs{\bm{r}_i(t)-\bm{r}_G(t)}$ 
is the unit vector
that gives the direction 
from the center of mass position of a cluster $\bm{r}_G=\sum_{i=1}^N\bm{r}_i/N$.
The probability distribution function of order parameters $\Phi(P,M)$ is defined
to satisfy the normalization condition
\begin{equation}
\label{}
1=\int_0^1\dd P\int_0^1\dd M\Phi(P,M),
\end{equation}
and we set the width of bin as $\Delta P=\Delta M=0.01$ in a simulation.
In addition, we measure the duration of the polar and rotational order $TP$ and $TM$ 
by the time that $P(t)$ and $M(t)$ exceed the threshold values $P_{th}=M_{th}=0.6$, respectively.

The radius of gyration is
\begin{equation}
\label{}
R_g(t)=\sqrt{\frac{1}{N}\sum_{i=1}^N\bm{c}_i^2(t)}.
\end{equation}
For example, in the case of the uniform disc with radius $R_0$,
$R_g$ is $R_0/\sqrt{2}$.
The moment of inertia tensor is 
\begin{equation}
\label{}
I_{\mu\nu}=\sum_{i=1}^N(\bm{c}_i^2\delta_{\mu\nu}-c_{i,\mu}c_{i,\nu}),
\end{equation}
where  $\mu,\nu=x,y$ and $\delta_{\mu\nu}$ is the Kronecker delta,
and the principal moments of inertia are defined as
$I_\alpha$ ($\alpha = 1,2$) to satisfy $I_1>I_2$.
The corresponding eigenvectors $\bm{u}_{\alpha}$ are obtained 
as $u_{\alpha,x}=I_{xy},~u_{\alpha,y}=I_\alpha-I_{xx}$ and then normalized as $\hat{\bm{u}}_{\alpha}$.
We define the direction-shape correlations $q_\alpha(t)$ as
\begin{equation}
\label{}
q_\alpha(t)=\abs{\bm{P}(t)\cdot\hat{\bm{u}}_\alpha(t)},
\end{equation}
where, due to $I_1>I_2$,
the direction of motion is perpendicular to the long axis of the cluster when $q_1\approx1$,
and is parallel to the long axis of the cluster when $q_2\approx1$.
In addition, we obtain the mean speed of agents:
\begin{equation}
\label{}
U(t)=\frac{1}{N}\sum_{i=1}^N\frac{v_i(t)}{v_0}.
\end{equation}
We define the probability of overlap $p_o(t)$ as
the ratio of the number of agents whose eye enters the body of a neighbor 
to the number of all agents $N$
at time $t$.

To construct the phase diagram, we performed 25 independent simulations for each parameter 
over the time interval $t\in[0,1000]$. 
The measured quantities are time-averaged in $t \in [300, 1000]$ and 
then ensemble-averaged across the 25 runs to obtain the averaged value.
We normalized $TP$ and $TM$ by the duration of averaging time $t\in[300,1000]$ so that they lie in [0,1].
Specifically, this average method is used in Fig.~\ref{Fig3}(a)-(d),
~\ref{FigE3}(b)-(f),~\ref{FigE4}(b)-(d), (f),~\ref{FigE5}, and Fig.~S1.

\subsubsection*{spatial distributions}
The normalized density distribution function $\rho(c)$ is 
a function of the radial distance $c$ from the center of mass of the cluster.
We define the density distribution $\widetilde{\rho}(c,t)$ as follows.
\begin{equation}
\label{}
\widetilde{\rho}(c,t)=\sum_{i=1}^N\frac{\delta(c_i(t)-c)}{c},
\end{equation}
where $c_i(t)$ is the radial distance of agent $i$, 
and $\delta(\circ)$ is the Dirac delta function. 
Numerically, we use the bin width $\Delta c=0.05$.
Then we calculate the normalized density distribution
\begin{equation}
\label{}
\rho(c,t)=\frac{\widetilde{\rho}(c,t)}{\int_0^\infty dc c\widetilde{\rho}(c,t)}.
\end{equation}

The distribution of the speed is defined as
\begin{equation}
\label{}
v(c,t)=\sum_{i=1}^Nv_i(t)\delta(c_i(t)-c)
\end{equation}
The time averaging is performed only at the time step when $\rho(c,t)>0$.
Therefore, $v(c)\rho(c)$ shows the distribution of expected speed.

In addition, we measure the probability distributions of the minimum distance $P(d,t)$:
\begin{equation}
\label{}
P(d,t)=\frac{1}{N}\sum_{i=1}^N\delta(d_i(t)-d),
\end{equation}
where $d_i(t)=\min_{j\neq i}r_{ij}(t)$ is the minimum distance between the eye positions of agents.

\subsubsection*{correlation function}
We define the mean velocity as
\begin{equation}
\label{}
\bm{V}(t)=\frac{1}{N}\sum_{i=1}^N\bm{v}_i(t),
\end{equation}
and the velocity fluctuation as $\delta\bm{v}_i(t)=\bm{v}_i(t)-\bm{V}(t)$.
The correlation function of velocity fluctuations is
\begin{equation}
\label{defC}
C(r)=\frac{\sum_{i,j=1}^N\delta\bm{v}_i(t)\cdot\delta\bm{v}_j(t)\delta(r-r_{ij}(t))}{C_0\sum_{i,j=1}^N\delta(r-r_{ij}(t))},
\end{equation}
where $C_0$ is the normalization constant for the condition $C(r=0)=1$,
and we use the bin width $\Delta r=0.05$.

Here, we define $B(r)=\sum_{i,j=1}^N\delta(r-r_{ij}(t))$,
and then Eq.~(\ref{defC}) is integrated with respect to $r$ as
\begin{equation}
\label{}
C_0\int_0^\infty\dd rC(r)B(r)=\sum_{i,j=1}^N\delta\bm{v}_i(t)\cdot\delta\bm{v}_j(t)
\end{equation}
due to $\int_0^\infty\dd r\delta(r-r_{ij}(t))=1$.
Thus, because $\sum_{i=1}^N\delta\bm{v}_i(t)=0$ and $C_0>0$,
we obtain $\int_0^\infty\dd rC(r)B(r)=0$,
and then find that
$C(r)$ has a distance (called the correlation length $\xi$) at which the correlation is zero ($C(\xi)=0$)
due to $B(r)>0$.
In other words, the change of motion of an agent is transmitted to neighbors in the range $r\lesssim\xi$.

For group sizes $N\gtrsim500$, cluster splitting occurs frequently, 
and it is therefore difficult to maintain a stable cluster over timescale $t\sim\mathcal{O}(10^4)$.
Thus, we did not measure $\xi$ for large clusters:
such splitting is commonly observed in 2D models~\cite{Calovi2014,Ito2024},
and a 3D model can reproduce a stable cluster for up to 10000 agents~\cite{Ito2022}.

\subsection*{Derivation of $\Theta_{\mathrm{rand}}(n;\chi_1,\chi_2)$}
In the case of randomly positioned agents,
$\Theta_n$ approximately becomes the product of
the horizontal angular diameter of the $n$NN $\delta_n^\parallel$
and the probability $P_n$ that the $n$NN remains visible through gaps without occlusion by the closer $(n-1)$NN.
When the spatial density of agents is uniform,
the distance to the $n$NN scales as $r_n \propto\sqrt{n}$:
$\delta_n^\parallel=2\tan[-1](l/2r_n)=2\tan[-1](\chi_2/\sqrt{n})$,
where $\chi_2$ is a constant.
Next, let $S(n)$ be the total angular size of the unoccluded gaps remaining after placing $n$NN 
(with the initial condition $S(0)=2\pi$).
By applying a continuous approximation in the region of large $n$, 
we obtain the following differential equation:
\begin{equation}
\label{}
\dv{S}{n}=-\frac{\delta_n^\parallel}{2\pi}S=-\frac{1}{\pi}\tan[-1](\frac{\chi_2}{\sqrt{n}})S.
\end{equation}
However, due to finite-size effects, agents near the boundary of a cluster 
perceive an anisotropic distribution of neighbors, 
which suppresses the occlusion probability compared to the ideal isotropic case. 
Thus, we replace $1/\pi$ with an effective screening coefficient $\chi_1<1/\pi$,
and then
\begin{equation}
\label{}
\dv{S}{n}=-\chi_1\tan[-1](\frac{\chi_2}{\sqrt{n}})S
\end{equation}
yields
\begin{eqnarray}
\label{}
P_n&=&\frac{S(n)}{2\pi}\\\nonumber
&\propto&\exp[-\chi_1\leri{(n+\chi_2^2)\tan[-1](\frac{\chi_2}{\sqrt{n}})+\chi_2\sqrt{n}}].
\end{eqnarray}
Therefore, $\Theta_{\mathrm{rand}}(n;\chi_1,\chi_2)=\delta_n^\parallel P_n$ 
as Eq.~(\ref{Thetarand}).

\subsection*{Optimization of acceleration}
We show the optimization method of Eq.~(\ref{funcacc}).
We discretize $r$, $\phi$, and $\psi$ by dividing each into bins, 
and label them by integers $k$, $l$, and $m$, respectively.
For example, we denote $a_{klm}=a(r_k,\phi_l,\psi_m)$ and $A_{1,k}=A_1(r_k)$ for simplicity.
Note that we adopt $r_k>0.5$
to exclude cases in which an agent's eye overlaps with another agent.

In a two-agent system, we get the time-averaged $a_{klm}$ 
over an interval of $t\in[10,50]$ within $t\in[0,50]$ for $10^6$ simulations.
As the initial condition, two agents are placed within a circle of radius 3
with random positions and direction of motion,
and both have speed $v_0$.
For the measured $a_{klm}$ in this way, 
we define the optimization function as follows:
\begin{equation}
\label{}
Q=\sum_{k,l,m}n_{klm}\leri{a_{klm}-A_{1,k}O_{1,l}E_{1,m}-A_{2,k}E_{2,l}O_{2,m}}^2,
\end{equation}
where $n_{klm}$ is the number of data samples in a bin indexed ($k,l,m$).
At each iteration step of optimization, we obtain $A_{1,k}$ and the other functions 
by setting their partial derivatives of $Q$ to zero, 
and substitute them into $Q$.
(See Supplementary Section I for the details of calculations of partial derivatives).
Iterations are continued until the condition of
\begin{equation}
\label{}
\frac{\abs{Q_{\mathrm{current~step}}-Q_{\mathrm{previous~step}}}}{Q_{\mathrm{previous~step}}}<\epsilon
\end{equation}
is satisfied,
where $\epsilon=10^{-5}$.
Note that the relative error $Q/(\sum_{k,l,m}n_{klm}a_{klm}^2)$ 
at the final iteration step is sufficiently small compared to 1
(e.g. 0.035 and 0.025 for the parameters of the vortex and the polarized school (L), respectively).

We set the width of bins $\Delta r=0.05$, $\Delta\phi=\Delta\psi=10^\circ$.
To stabilize the optimization, 
we introduce the following normalization conditions on the angular functions 
$O_1(\phi), E_1(\psi), E_2(\phi), O_2(\psi)$~\cite{Calovi2018}:
\begin{equation}
\label{}
\frac{1}{2\pi}\int_{-\pi}^\pi\dd\phi O_1^2(\phi)=1,
\end{equation}
with analogous conditions for the other functions.
We absorb the normalization constants obtained in this way into $A_1(r),A_2(r)$.
In addition, for each distance bin $r_k$, 
there are $36\times36=1296$ corresponding angular bins $\phi_l,~\psi_m$. 
Therefore, only acceleration data in distance bin $r_k$ with $\sum_{lm}n_{klm}>129600$ samples 
is considered valid and displayed in Fig.~\ref{Fig5} and~\ref{FigE6},
because each bin contains an average of more than 100 samples across angular bins.

\newpage
\begin{figure*}[!t]
\centering
\includegraphics[width=\linewidth]{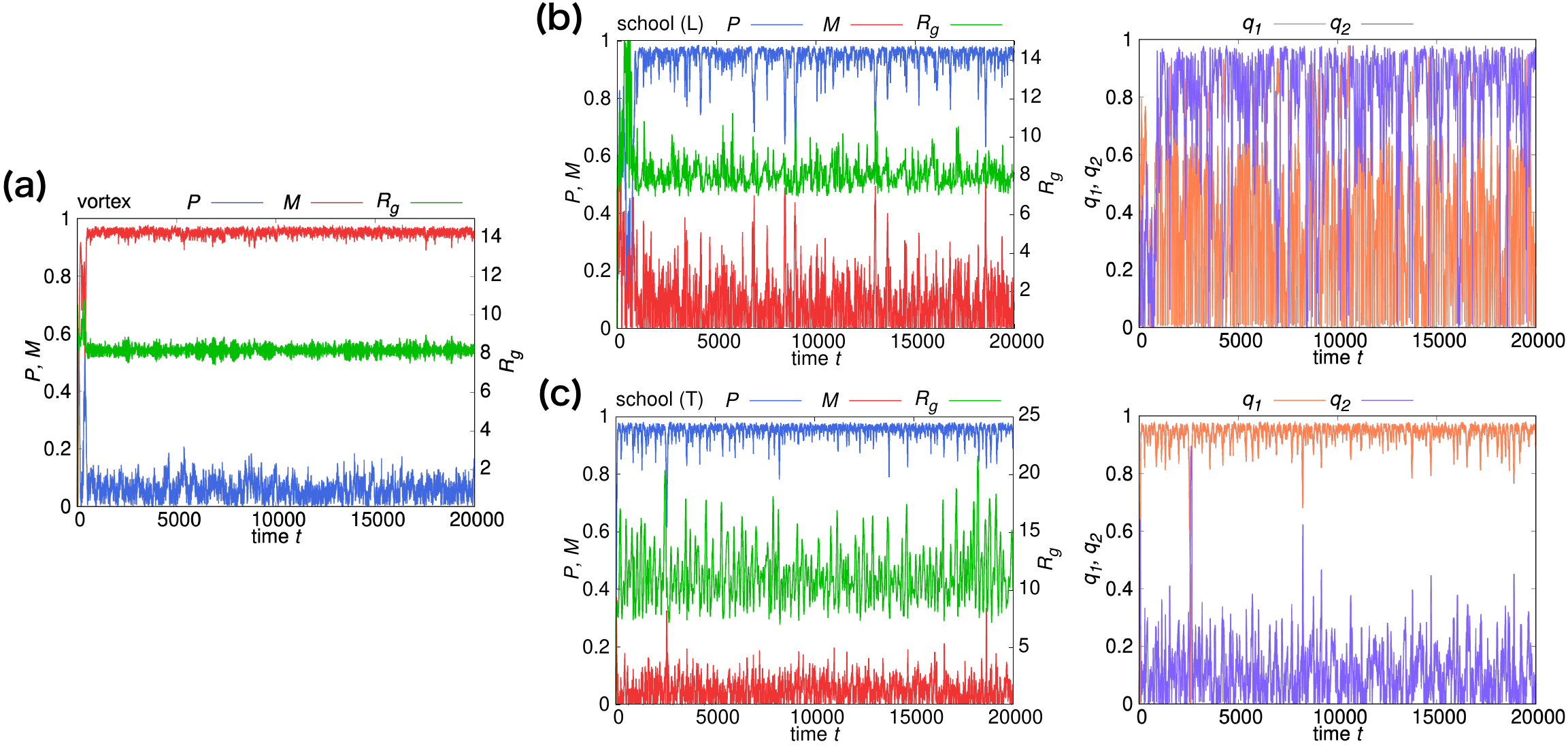}
\caption{
Time evolution for (a) vortex, (b) polarized school (L), and (c) polarized school (T).
The time window is $t\in[0,20000]$.
$P$ is polar order parameter (blue),
$M$ is rotational order parameter (red), 
$R_g$ is radius of gyration (green),
and $q_1,~q_2$ are direction-shape correlations (orange and purple, respectively).
In polarized school (L), the state where $q_2>q_1$ persists for most of the time,
indicating that the cluster is elongated parallel to the direction of motion.
In polarized school (T), the state where $q_1>q_2$ persists for most of the time,
indicating that the cluster is elongated perpendicular to the direction of motion.
}
\label{FigE1}
\end{figure*}

\begin{figure*}[!t]
\centering
\includegraphics[width=\linewidth]{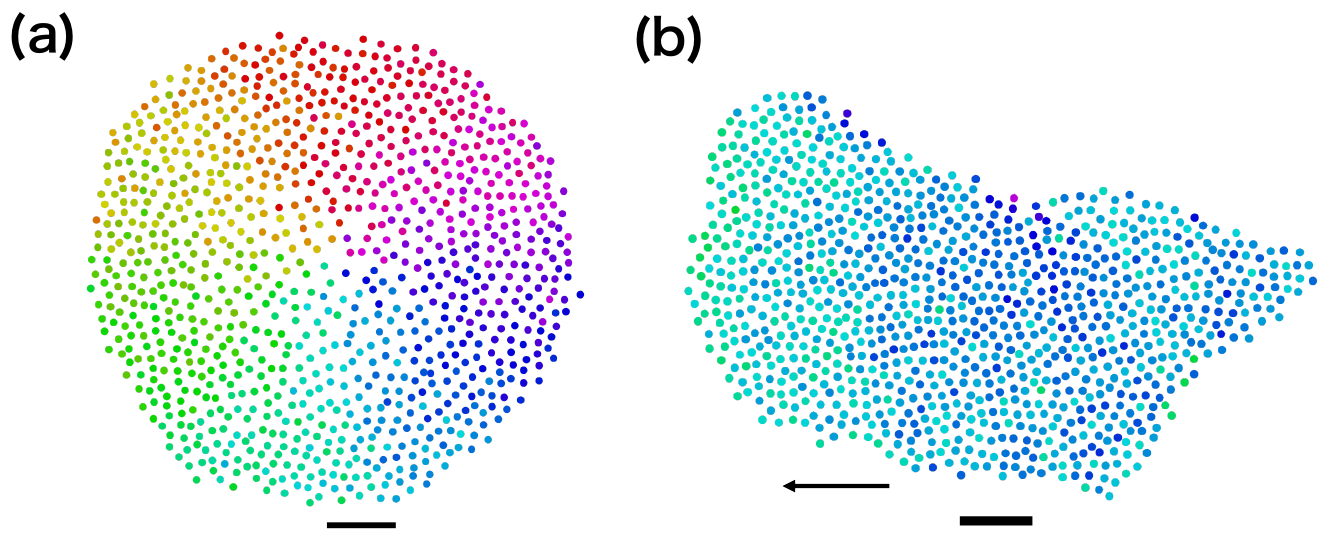}
\caption{
Snapshots of (a) vortex and (b) polarized school with $N=1000$.
The parameters of (a) and (b) are the same as those in Fig.~\ref{Fig2} (c) vortex and (d) polarized school, respectively.
The display of the agent and the scale bar are the same as those in Fig.~\ref{Fig2}(a)-(e).
See Supplemental material Movies S7-S8 for the dynamics.
}
\label{FigE2}
\end{figure*}

\begin{figure*}[!t]
\centering
\includegraphics[width=\linewidth]{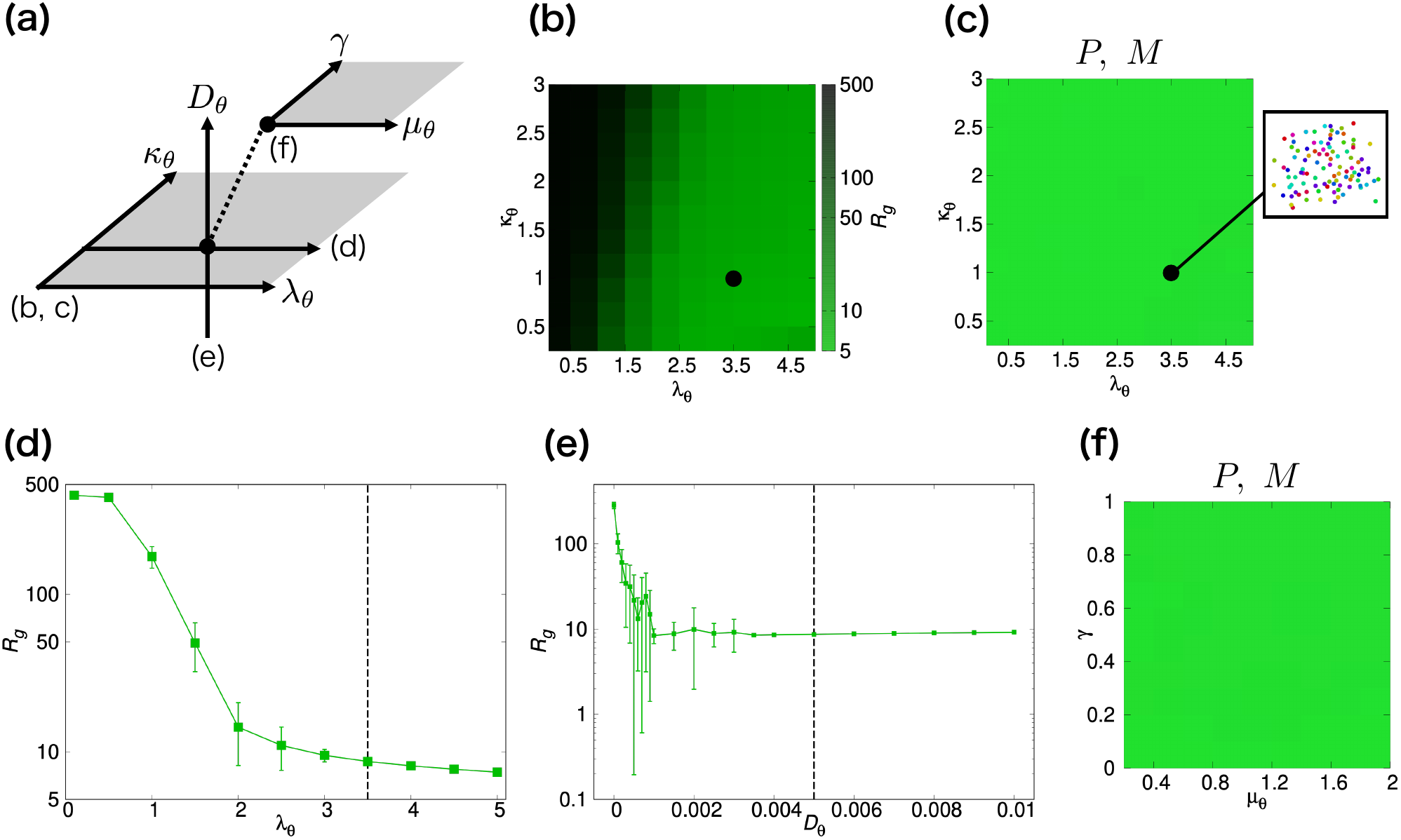}
\caption{
Cohesion of the cluster and emergence of swarm with $N=100$.
(a) The schematic figure shows the parameter regions of (b)-(f).
The other parameters $\kappa_v$, $\lambda_v$, $\mu_v$, $D_v$ are fixed to zero.
The black point represents 
$\kappa_\theta=1.0$, $\lambda_\theta=3.5$, $D_\theta=0.005$, $\mu_\theta=0$, $\gamma=0$
at which the swarm emerges (see Fig.~\ref{Fig2}(a)).
(b) The radius of gyration $R_g$ and (c) the order parameters $P,M$
where the horizontal axis is $\lambda_\theta$, and the vertical axis is $\kappa_\theta$.
(d) Dependence of $R_g$ on $\lambda_\theta$ (the axis along $\kappa_\theta=1.0$ in (b)).
The vertical dashed line represents $\lambda_\theta=3.5$.
(e) Dependence of $R_g$ on $D_\theta$.
The vertical dashed line represents $D_\theta=0.005$.
In (d)-(e), the error bars represent the standard deviation over 25 simulations 
for the time-averaged values at $t\in[300,1000]$.
(f) Dependence of $P,M$ on $\mu_\theta,~\gamma$.
In (c) and (f), the color represents the value of order parameters as shown in Fig.~\ref{Fig3}(e).
}
\label{FigE3}
\end{figure*}

\begin{figure*}[!t]
\centering
\includegraphics[width=\linewidth]{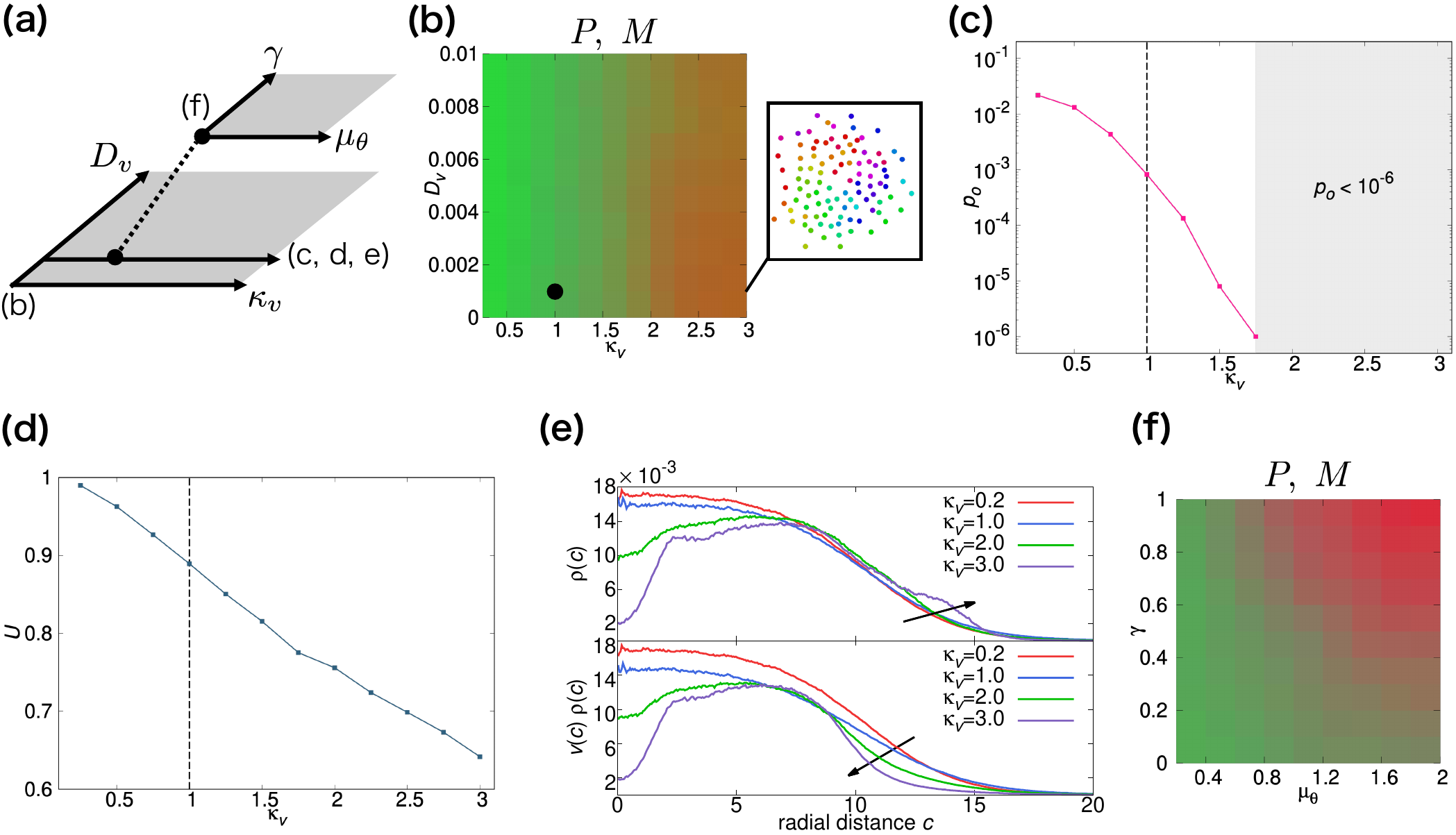}
\caption{
Resolving overlap and emergence of shell with $N=100$.
(a) The schematic figure shows the parameter regions of (b)-(f).
The other parameters $\lambda_v$ and $\mu_v$ are fixed to zero,
and $\kappa_\theta=1.0$, $\lambda_\theta=3.5$, $D_\theta=0.005$.
The black point represents 
$\kappa_v=1.0$, $D_v=0.001$, $\mu_\theta=0$, $\gamma=0$.
(b) Dependence of $P,M$ on $\kappa_v,~D_v$.
The shell emerges at $\kappa_v=3.0$, $D_v=0.001$ (see Fig.~\ref{Fig2}(b)).
Dependence of (c) the probability of overlap $p_o$ and (d) the mean speed $U$
on $\kappa_v$
(the axis along $D_v=0.001$ in (b)).
These are obtained by the average over 25 simulations.
The vertical dashed line represents $\kappa_v=1.0$.
In (c), $p_o$ is less than $10^{-6}$ at $\kappa_v>1.75$.
(e) The normalized density distribution $\rho(c)$ (upper figure)
and the distribution of expected speed $v(c)\rho(c)$ (lower figure)
as a function of radial distance from the center of mass $c$.
These are averaged over $t\in[300,20000]$ in a simulation.
$\kappa_v=0.2$: red, $\kappa_v=1.0$: blue, $\kappa_v=2.0$: green, $\kappa_v=3.0$: purple.
The black arrows show that, at $r\sim10$-15, $\rho(c)$ increases and $v(c)\rho(c)$ decreases
as $\kappa_v$ increases.
See also the Methods section ``spatial distributions" for the details of definition of $\rho(c),~v(c)$.
(f) Dependence of $P,M$ on $\mu_\theta,~\gamma$.
In (b) and (f), the color represents the value of order parameters as shown in Fig.~\ref{Fig3}(e).
}
\label{FigE4}
\end{figure*}

\begin{figure*}[!t]
\centering
\includegraphics[width=\linewidth]{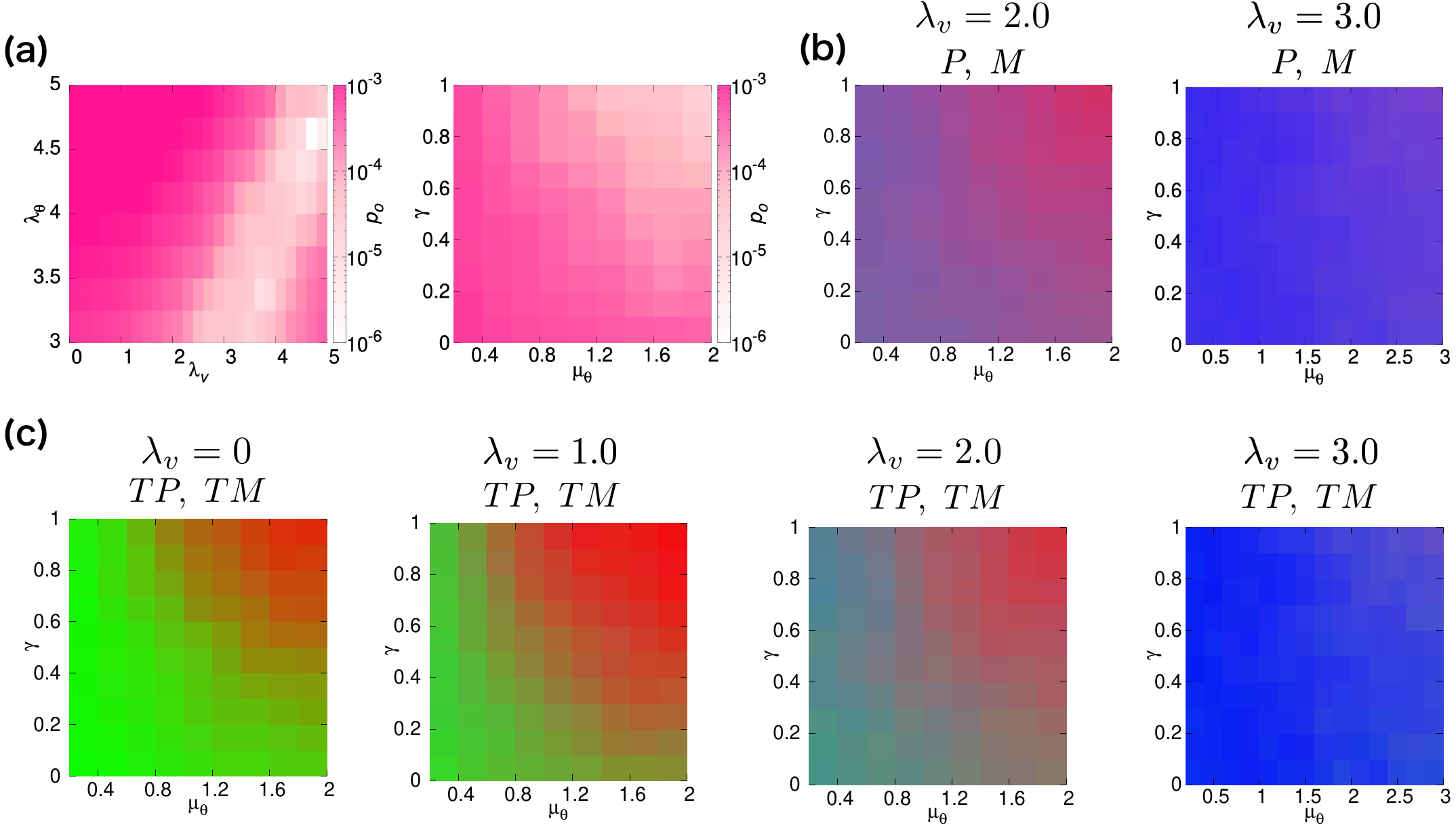}
\caption{
Additional phase diagrams with $N=100$.
$\kappa_\theta=1.0$, $\kappa_v=1.0$, $D_v=0.001$, $D_\theta=0.005$, and $\mu_v=0$ are fixed.
(a) The probability of overlap $p_o$ as a function of $\lambda_v,\lambda_\theta$ (left, $\mu_\theta=\gamma=0$)
and $\mu_\theta,\gamma$ (right, $\lambda_v=1.0$ and $\lambda_\theta=3.5$).
The parameter regions are the same as Fig.~\ref{Fig3}(a) and Fig.~\ref{Fig3}(b), respectively.
(b) Dependence of the order parameters $P,M$ on $\mu_\theta,\gamma$
with $\lambda_v=2.0$ (left) and $\lambda_v=3.0$ (right), and $\lambda_\theta=3.5$.
The color represents the value of order parameters as shown in Fig.~\ref{Fig3}(e).
Note that~\ref{FigE4}(f), Fig.~\ref{Fig3}(b), and figure (b) in this figure
correspond to the cases where $\lambda_v$ is increased from 0 to 3.0 in increments of 1.0.
(c) Dependence of the duration of orders $TP,TM$ on $\mu_\theta,\gamma$
with $\lambda_\theta=3.5$ and from left to right, $\lambda_v=0,~1.0,~2.0,~3.0$, respectively.
Note that these figures show $TP,TM$ over the same parameter range 
as that used for~\ref{FigE4}(f), Fig.~\ref{Fig3}(b), and figure (b) in this figure.
The color represents the value of $TP,TM$ as shown in Fig.~\ref{Fig3}(e)
by replacing $P$ and $M$ with $TP$ and $TM$, respectively.
}
\label{FigE5}
\end{figure*}

\begin{figure*}[!t]
\centering
\includegraphics[width=\linewidth]{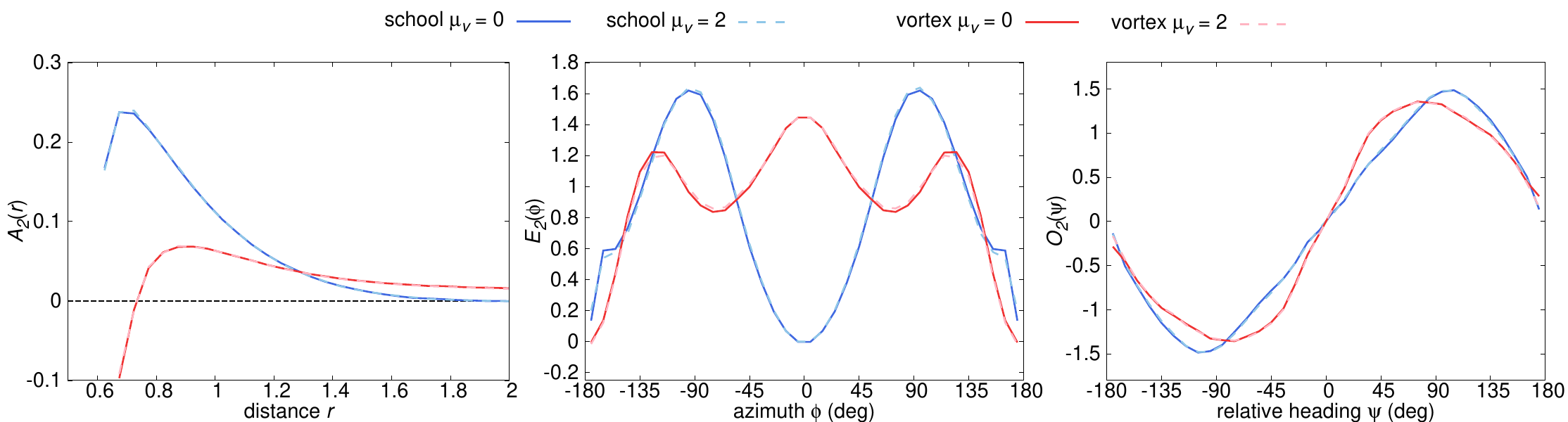}
\caption{
Dependence of alignment term on $\mu_v=0$ and 2.0
for the vortex ($\lambda_v=1.0$, $\mu_\theta=2.0$, $\gamma=0$)
and the school ($\lambda_v=3.0$, $\mu_\theta=\gamma=0$).
The solid lines and the dashed lines represent the case of $\mu_v=0$ and $\mu_v=2.0$, respectively.
$A_2(r)$ is left, $O_2(\phi)$ is center, and $E_2(\psi)$ is right.
}
\label{FigE6}
\end{figure*}

\begin{figure*}[!t]
\centering
\includegraphics[width=\linewidth]{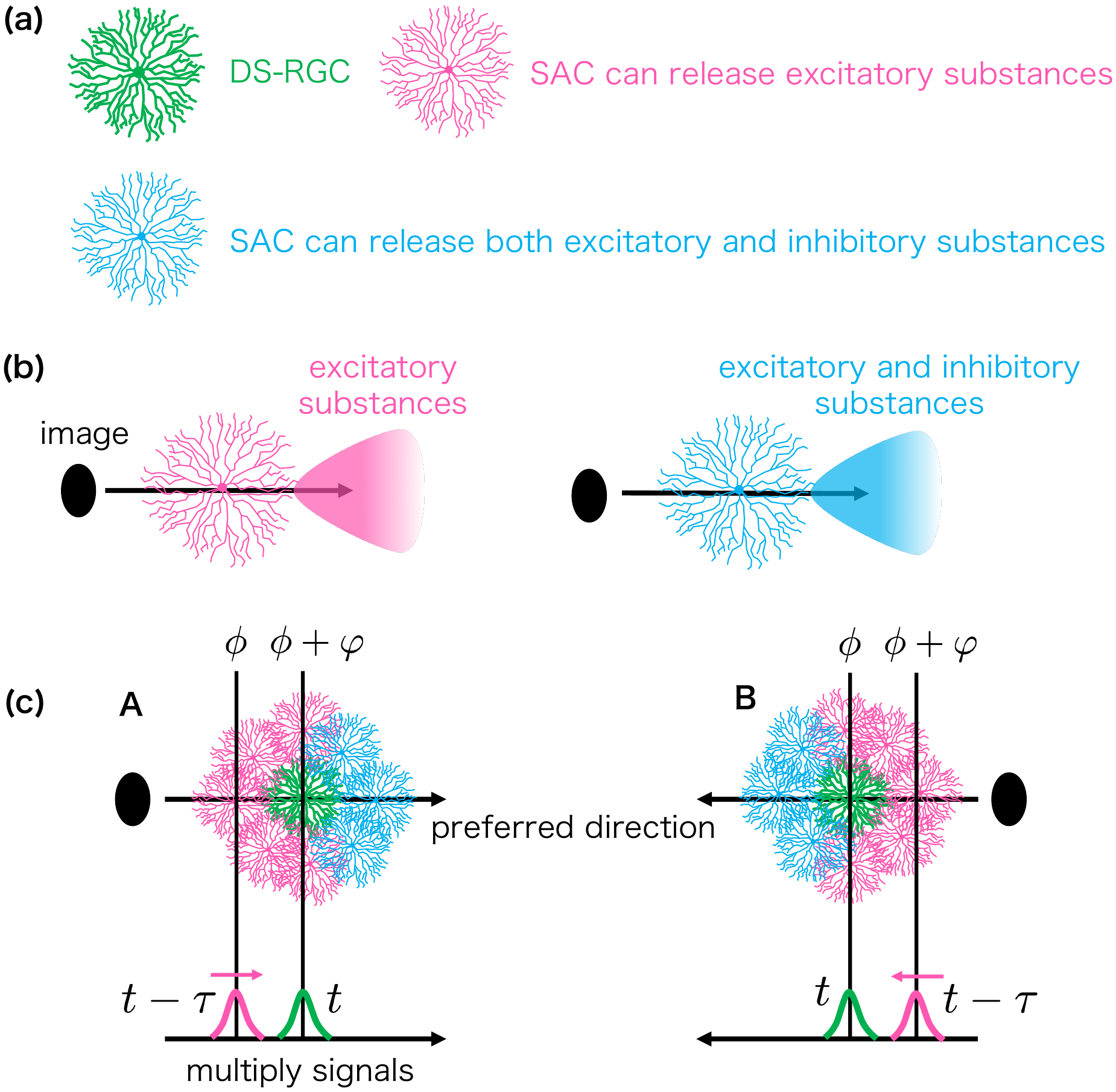}
\caption{
The mechanism of DS-RGC.
(a) The types of cells associated with optic flow detection in the retina.
(b) The dendritic branch responds to the passing image from the center toward the periphery,
and substances are released from the side on this branch.
(c) The asymmetrical connections of a DS-RGC and SACs.
The delayed signal (pink) from the SAC is multiplied 
with the instantaneous signal (green) on DS-RGC.
A: the positive preferred direction case.
B: the negative preferred direction case.
}
\label{FigE7}
\end{figure*}



\end{document}